\documentstyle[preprint,aps,epsf]{revtex}
\newcommand{\singlefig}[2]{
\begin{center}
\begin{minipage}{#1}
\epsfxsize=#1
\epsffile{#2}
\end{minipage}
\end{center}}
\newcommand{\segmentfig}[3]{
\begin{minipage}{#1}
\epsfxsize=#1
\epsffile{#2}
\begin{center}
{\small \mbox{#3}}
\end{center}
\end{minipage}}
\begin{document}
\draft
\title{Non-Abelian Black Holes in Brans-Dicke Theory}
\author{Takashi Tamaki\thanks{electronic
mail:tamaki@gravity.phys.waseda.ac.jp}
 and Kei-ichi Maeda\thanks{electronic
mail:maeda@gravity.phys.waseda.ac.jp}
 }
\address{Department of Physics, Waseda University,
Shinjuku, Tokyo 169, Japan}
\author{Takashi Torii\thanks{electronic
mail:torii@th.phys.titech.ac.jp}
 }
\address{Department of Physics, Tokyo Institute of
Technology, Meguro, Tokyo 152, Japan}
\date{\today}
\maketitle
\begin{abstract}
\baselineskip=.5cm
We find a black hole solution with non-Abelian field in Brans-Dicke theory. It is an extension of non-Abelian black hole in general
relativity. We discuss two non-Abelian fields: ``SU(2)" Yang-Mills
field with a mass (Proca field) and the SU(2)$\times$SU(2) Skyrme field.  In both cases, as in general relativity, there are two branches of
solutions, i.e., two black hole solutions with  the same horizon
radius.   Masses of both black holes are always smaller than
those  in general relativity.
A cusp structure in the mass-horizon radius
($M_{g}$-$r_{h}$) diagram, which is a typical symptom of
stability change in catastrophe theory, does not appear in the
Brans-Dicke frame but  is found in the Einstein conformal frame.
This suggests that catastrophe theory may be simply applied for a
stability analysis as it is if we  use the variables in the
Einstein frame. We also discuss the
effects of the  Brans-Dicke  scalar field on black hole
structure.
\end{abstract}

\baselineskip=.8cm
\section{Introduction}                            %
 \ \ For many years, there have been various efforts to find a
theory of ``everything". The Kaluza-Klein  theory
was one of the candidates, which is constructed in a five  dimensional
spacetime. Jordan noticed in 1955 that in our four dimensional
spacetime,   a scalar field appears by a  compactification in the
Kaluza-Klein theory and it gives a nonminimal
coupling to gravity, meaning that this theory violates even
the weak equivalence principle. Dicke thought that the weak
equivalence principle must be guaranteed based on several experiments.
Then,  from the weak equivalence principle\footnote{But  for self-gravitating bodies, the weak equivalence principle is still violated
(Nordtvedt effect) \cite{rf:Nordtverdt}.} and Mach's Principle,
which insists  that an inertial force is determined by the
distribution of  matter all over the Universe,  he  and Brans
constructed  a new scalar-tensor theory, i.e., Brans-Dicke (BD) theory, in  1961\cite{rf:BD}.   Since then the
difference between general  relativity (GR) and BD theory has
been discussed in many aspects. Although BD  theory itself is
strongly constrained by several experiments  (the BD parameter
$\omega\geq 500$),  we believe that the theory may still be
important from the following points of view:
\begin{enumerate}
\item[] BD theory can be an effective field theory of a unified
theory of fundamental  forces. In particular
 the BD-type scalar field appears as  a dilaton field in superstring theory.
\item[]  BD theory is one of the simplest extensions of GR. So if
we wish to discuss  something in a generalized theory of gravity,
 BD theory can be the best model to  see a difference from GR.
\end{enumerate}
Moreover, a scalar field such as the BD scalar field may have an affect on
many aspects in gravitational physics.    For  example, the
inflationary scenario would be modified by an introduction of such
a scalar field\cite{rf:Linde}. Although the inflationary scenario
was discussed originally in GR, since an appropriate model based on
 particle physics has not been found, it is  important to
recognize that the introduction of a scalar field can make a
big  change in scenarios of the very early universe.

 Black holes are also important in gravitational physics.  We may expect that such a scalar field also affects some feathers  of a black hole\cite{GM}.  However, since the gravity part in  BD theory is conformally equivalent to that in GR, black hole solutions are not modified by the
introduction of the BD scalar field for the case without matter or
with a traceless matter field such as the electromagnetic field. As
a result, for vacuum case or the case with the electromagnetic
field, a  conventional  Kerr or Kerr-Newman black hole  turns out
to be a unique solution even in BD theory because of the
black hole no-hair theorem in GR\cite{rf:NH}. Hence, here we shall
discuss  a non-Abelian black hole  in BD theory, which has
 so far not been studied so much.  For the case with the Yang-Mills
field, however, we again find the same colored black hole as that
in GR\cite{rf:colored}, because its energy-momentum tensor is traceless.
Then,  we discuss a ``massive" non-Abelian field,
i.e., a massive Yang-Mills (Proca) field, and the Skyrme field.
We consider only the globally neutral case in this paper.

After the introduction of basic ans\"{a}tze and conditions in \S.II, we
present the Proca black hole solution and its properties in
\S.III. We find some difficulty in adopting catastrophe theory to the stability analysis.  To resolve such a difficulty, we
introduce new variables defined in the Einstein conformal frame in
\S.IV.  We find  quite similar properties of black hole solutions
to those in GR: in particular a cusp structure appears in the
mass-horizon radius diagram.  This allows simple  application of catastrophe theory in the stability analysis as it is. The effects
of the BD scalar field on black hole structure are investigated
in \S.V. In
\S.VI, we discuss a Skyrme black hole, showing that its properties
are quite similar to those in the Proca black hole. The concluding
remarks will follow in
\S.VII. Throughout this paper we use  units of
c=$\hbar$=1.  Notations and definitions such as Christoffel
symbols and curvatures follow Misner-Thorne-Wheeler
\cite{rf:MTW}.
\newpage
\section{Non-Abelian black holes in Brans-Dicke theory}    %
\ The action of BD theory is written as

\begin{eqnarray}
 S=\int d^{4}x\sqrt{-g}\left[\frac{1}{2\kappa^{2}}\left(\phi R-
\frac{\omega}
{\phi}\nabla_{\alpha}\phi\nabla^{\alpha}\phi
\right) + L_{m}\right]\ ,
\label{eq:2}
\end{eqnarray}
where $\kappa^{2} \equiv 8\pi G$ with $G$ being Newton's
gravitational constant. The BD parameter is $\omega$, and
$L_{m}$ is the Lagrangian of the matter field. The
dimensionless BD scalar field $\phi$ is normalized by $G$.

 For the BD field $\phi$, the field equation becomes
\begin{eqnarray}
  \Box \phi = \frac{\kappa^{2}}{2\omega+3}T^{\mu}_{\mu}\ .
\label{eq:1}
\end{eqnarray}
Then, if the right hand side of this equation vanishes, that is, the
energy-momentum tensor of the matter field is traceless, $\phi$= constant turns out to be  a solution, meaning that a black hole solution in
GR is also a solution in BD theory.
Hence, for the SU(2) Yang-Mills field, we find that the colored
black hole\cite{rf:colored} is a solution in BD theory too.   Although we have no proof, we expect that  for the case with a
massless non-Abelian gauge field, no new type of black
hole solution appears in BD theory.

If a non-Abelian field is massive  or effectively massive, however,
$\phi$ = constant is no longer a solution.  We will find a new type
of black hole solution, and can discuss some differences from 
black hole solutions in GR. This is  the reason for us to study a
massive non-Abelian field here.

We assume that a black hole is  static and spherically  symmetric, in
which case the metric is written as

\begin{eqnarray}
ds^{2}=-\left[1-\frac{2Gm(r)}{r}\right]e^{-2\delta (r)}dt^{2}+
\left[1-\frac{2Gm(r)}{r}\right]^{-1}dr^{2}+r^{2}d\Omega^{2}\ .
\label{eq:2.1}
\end{eqnarray}

The boundary condition for a black hole solution at spatial infinity
is\footnote{This choice of $\phi_{0}$ at $\infty$ guarantees that $G$
is Newton's gravitational constant\cite{rf:BD}. }
\begin{eqnarray}
\lim_{r\rightarrow \infty} m = M< \infty,\
\lim_{r\rightarrow \infty}\delta =0,\
\lim_{r\rightarrow \infty}\phi =
\phi_{0}\equiv
\frac{2(2+\omega)}{3+2\omega} \ \label{eq:2.2}.
\end{eqnarray}
Note that a test particle far from a black hole does not
move under an influence of  this ``mass" $M$, but feels a
gravitational attractive force given by a gravitational mass $M_g$.
$M_g$ is defined from the asymptotic behavior of the time-time component of the metric and given as
\begin{eqnarray}
M_g = M+ {1 \over G}\lim_{r \rightarrow \infty}(r\delta)\ .
\label{eq:2.x2}
\end{eqnarray}

For the existence of a regular event horizon, $r_h$, we have
\begin{eqnarray}
m_{h}\equiv m(r_h)=\frac{r_{h}}{2G},\
\delta_{h}\equiv\delta(r_h)<\infty\ .  \label{eq:12}
\end{eqnarray}
We also require that no singularity exists outside the
horizon, i.e.,
\begin{eqnarray}
m(r)<\frac{r}{2G}\  ~~~~~{\rm  for} ~~~ r>r_{h}. \label{eq:15}
\end{eqnarray}

For our numerical calculation, we introduce the following
dimensionless variables:
\begin{eqnarray}
\bar{r}=r/r_{h},\ \bar{m}=Gm/r_{h}.
\label{eq:2.3}
\end{eqnarray}

To write down the explicit equations of motion, we  have to specify
our models.  In what follows, we discuss the Proca field and
the Skyrme field, separately.
\section{Proca Black Hole}   %
\ \ We first consider a massive ``SU(2)" Yang-Mills  field (Proca
field).  The matter Lagrangian  $L_{m}$  is now
\begin{eqnarray}
 L_{m}=-\frac{1}{16\pi g_{c}^{2}}{\rm Tr}\mbox{\boldmath $F$}^{2}-
   \frac{\mu^{2}}{8\pi g_{c}^{2}}{\rm Tr}\mbox{\boldmath $A$}^{2}\ ,
\label{eq:p1}
\end{eqnarray}
where $g_{c}$ and $\mu$ are the gauge coupling constant and the
mass of the Proca field, respectively.
$\mbox{\boldmath $F$}$ is the field strength expressed by its
potential
$\mbox{\boldmath $A$}$ as
$\mbox{\boldmath $F$}=d\mbox{\boldmath $A$} +\mbox{\boldmath
$A$}\wedge\mbox{\boldmath $A$}$. For the spherically symmetric case,
we can write the vector potential as
\begin{eqnarray}
\mbox{\boldmath $A$}=a(r,t)\mbox{\boldmath $\tau$}_{r}dt
+\frac{b(r,t)}{r}\mbox{\boldmath $\tau$}_{r}dr
+\{d(r,t)\mbox{\boldmath
$\tau$}_{\theta}-[1+w(r,t)]\mbox{\boldmath $\tau$}_{\phi}\}d\theta
\nonumber \\
+\{[1+w(r,t)]\mbox{\boldmath $\tau$}_{\theta}+d(r,t) \mbox{\boldmath
$\tau$}_{\phi}\}\sin \theta d\phi
\label{eq:p2}
\end{eqnarray}
as Witten showed\cite{rf:Witten}, where $\mbox{\boldmath
$\tau$}_{r}, \mbox{\boldmath $\tau$}_{\theta}$,
and $\mbox{\boldmath$\tau$}_{\phi}$ are the generators of su(2) Lie algebra. We adopt the 't Hooft ansatz, i.e., $a\equiv 0$, which means that  only a magnetic component of the Proca field exists. We also set $b=0$.\footnote{If the Yang-Mills
field is massless, we can always  impose this
condition via a gauge transformation.  In  the present case,
however, we can just show that  this is consistent with the field
equation.} In the static case, we can set
$d=0$. Now, our potential is
\begin{eqnarray}
\mbox{\boldmath $A$}=[1+w(r)](-\mbox{\boldmath
$\tau$}_{\phi}d\theta+
 \mbox{\boldmath $\tau$}_{\theta}\sin\theta d\phi) \ .
\label{eq:p3}
\end{eqnarray}
The boundary condition of the Proca field for its total energy to
be finite  is
\begin{eqnarray}
\lim_{r\rightarrow \infty} w = -1\ . \label{eq:p4}
\end{eqnarray}
We define dimensionless parameters as
\begin{eqnarray}
\bar{\mu}=\mu/(g_{c}m_{p}),\ \ \lambda_{h}=r_{h}/(l_{p}/g_{c}) \ .
\label{eq:6}
\end{eqnarray}
 $l_{p} \equiv G^{1/2}$ and $m_{p} \equiv G^{-1/2}$ are the Planck
length and mass defined by Newton's gravitational constant, respectively.

Under the above ans\"{a}tze,  we find the following basic
equations:
\begin{eqnarray}
\frac{d\bar{m}}{d\bar{r}}&=&\left(2\phi+\bar{r}\frac{d\phi}{d\bar{r}} \right)^{-1}
\left[ \left(1-\frac{2\bar{m}}{\bar{r}}\right)
\left\{
\frac{\omega+2}{2\phi} \left(\bar{r}\frac{d\phi}{d\bar{r}}\right)^{2}
+ \frac{2}{\lambda_{h}^{2}} \left(\frac{dw}{d\bar{r}}\right)^{2}
\right\}
-\bar{m}\frac{d\phi}{d\bar{r}}   \right.  \nonumber   \\
&&+\left.\frac{1}{\lambda_{h}^{2}}\left( \frac{1-w^{2}}{\bar{r}}
\right)^{2}
 \left(1+\frac{\bar{r}}{\phi}\frac{d\phi}{d\bar{r}} \right)
+2(1+w)^{2}\bar{\mu}^{2}
\left\{
\frac{2\omega+1}{2\omega+3}
+\frac{2(\omega+1)\bar{r}}{(2\omega+3)\phi}
\frac{d\phi}{d\bar{r}}
\right\}\right]\ ,          \label{eq:7}           \\
\frac{d\delta}{d\bar{r}}&=&-\frac{\bar{r}}{\phi}
\left( 2\phi+\bar{r}\frac{d\phi}{d\bar{r}} \right)^{-1}
\left\{
(\omega+1)\left(\frac{d\phi}{d\bar{r}}\right)^{2}
+\frac{4}{\lambda_{h}^{2}}\frac{\phi}{\bar{r}^{2}}
\left(\frac{dw}{d\bar{r}}\right)^{2}
\right\}            \nonumber        \\
&&-\bar{r}\left(2\phi+\bar{r}\frac{d\phi}{d\bar{r}}\right)^{-1}
\left(1-\frac{2\bar{m}}{\bar{r}}\right)^{-1}
\left[
\frac{1}{\lambda_{h}^{2}}\frac{\bar{r}}{\phi}\frac{d\phi}{d\bar{r}}
\left\{
\left(\frac{1-w^{2}}{\bar{r}^{2}}\right)^{2}
+\frac{4(\omega+1)\bar{\mu}^{2}\lambda_{h}^{2}}{2\omega+3}
\left(\frac{1+w}{\bar{r}}\right)^{2}
\right\}\right.     \nonumber \\
&&-\left.\frac{2}{\bar{r}}
\left(1-\frac{\bar{m}}{\bar{r}}\right)\frac{d\phi}{d\bar{r}}
-\frac{4\bar{\mu}^{2}}{2\omega+3}\left(\frac{1+w}{\bar{r}}\right)^{2}
\right]\ ,     \label{eq:8}       \\
\frac{d^{2}\phi}{d\bar{r}^{2}}&=&\frac{1}{\phi}\left(
\frac{d\phi}{d\bar{r}}\right)^{2}
+\left(1-\frac{2\bar{m}}{\bar{r}}\right)^{-1}
\left[
\frac{1}{\lambda_{h}^{2}}\frac{\bar{r}}{\phi}\frac{d\phi}{d\bar{r}}
\left\{
\left( \frac{1-w^{2}}{\bar{r}^{2}} \right)^{2}
+\frac{4(\omega+1)\bar{\mu}^{2}\lambda_{h}^{2}}{2\omega+3}
\left( \frac{1+w}{\bar{r}} \right)^{2}
\right\}\right.     \nonumber \\
&&-\left.\frac{2}{\bar{r}}\left(1-\frac{\bar{m}}{\bar{r}} \right)
\frac{d\phi}{d\bar{r}}
-\frac{4\bar{\mu}^{2}}{2\omega+3}\left(\frac{1+w}{\bar{r}}\right)^{2}
\right]\ ,   \label{eq:9}     \\
\frac{d^{2}w}{d\bar{r}^{2}}&=&\frac{1}{\phi}\frac{dw}{d\bar{r}}
\frac{d\phi}{d\bar{r}} +\left(1-\frac{2\bar{m}}{\bar{r}}\right)^{-1}
\left[
(1+w)\left\{\bar{\mu}^{2}
\lambda_{h}^{2}-\frac{w(1-w)}{\bar{r}^{2}}\right\}
-\frac{2\bar{m}}{\bar{r}^{2}}\frac{dw}{d\bar{r}}
\right.
\nonumber      \\
&&+\left. \frac{1}{\lambda_{h}^{2}}\frac{\bar{r}}{\phi}\frac{dw}{d\bar{r}}\left\{
\left(\frac{1-w^{2}}{\bar{r}^{2}}\right)^{2}
+\frac{4(\omega+1)\bar{\mu}^{2}\lambda_{h}^{2}}{(2\omega+3)}
\left(\frac{1+w}{\bar{r}}\right)^{2}
\right\}
\right]\ .
\label{eq:10}
\end{eqnarray}

As for the boundary condition at the event horizon, in order for the
horizon to be regular, the square brackets in
(\ref{eq:8}), (\ref{eq:9}) and (\ref{eq:10}) must vanish at the
horizon
$\bar{r}=1$. Hence, we find that
\begin{eqnarray}
\frac{dw}{d\bar{r}}\bigg|_{\bar{r}=1}&=&
-\frac{
\phi_{h}\lambda_{h}^{2}(2\omega+3)(1+w_{h})  \{
\bar{\mu}^{2}\lambda_{h}^{2}- w_{h}(1-w_{h}) \}
}
{
(2\omega+3)\{ (1-w^{2}_{h})^{2}
-\phi_{h}\lambda_{h}^{2} \}
+4\bar{\mu}^{2}(\omega+1)\lambda_{h}^{2}
(1+w_{h})^{2}
}\ ,      \label{eq:13}   \\
\frac{d\phi}{d\bar{r}}\bigg|_{\bar{r}=1}&=&
\frac{
4\bar{\mu}^{2}
(1+w_{h})^{2}\lambda_{h}^{2}\phi_{h}
}
{
(2\omega+3)\{ (1-w^{2}_{h})^{2}
-\phi_{h}\lambda_{h}^{2} \}
+4\bar{\mu}^{2}(\omega+1)\lambda_{h}^{2}
(1+w_{h})^{2}
}\ ,      \label{eq:14}
\end{eqnarray}
where $w_{h}\equiv w(r_h)$ and $\phi_{h}\equiv \phi(r_h)$. As a
result, $w_{h}$  and
$\phi_{h}$ turn out to be  shooting parameters and should be
determined iteratively so that the boundary conditions
(\ref{eq:2.2}) and (\ref{eq:p4}) are satisfied.

In Fig.1, we present a numerical solution with
$r_{h} = 0.5 l_p/g_c$ and $\mu = 0.15 g_c m_p$.\footnote{As we see later, for fixing $r_{h}$, we have two solutions in BD theory as in GR. Here, we show the field distributions in the solid-line branch 
in Fig.\protect\ref{m-rm0.15}.} We set $\omega=0$. Although
this is not consistent with the present limit from experiments, we
choose this value because we wish to clarify the difference from 
black holes in GR.  For a massive non-Abelian field, the node number
of the potential $w(r)$ is limited by some finite integer. Here the node
number is chosen to be the smallest value, i.e., one.
The dotted line denotes the Proca black hole in GR\cite{rf:Greene} with the same parameters, i.e.
$r_{h} = 0.5 l_{p}/g_{c}$ and $\mu = 0.15 g_{c} m_{p}$, which we show as
a reference.

\begin{figure}[htbp]
\begin{tabular}{ll}
\segmentfig{8cm}{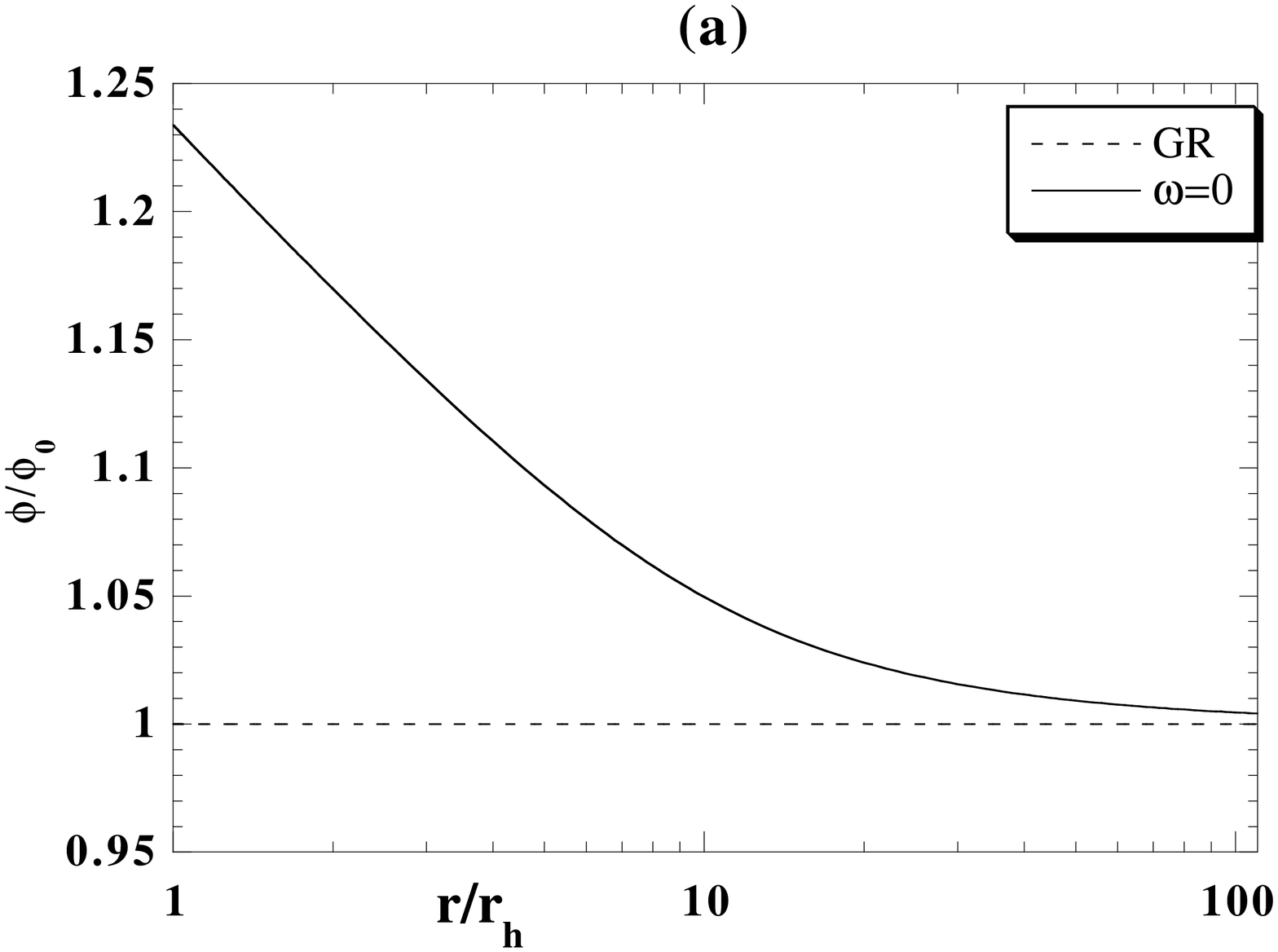}{}
\segmentfig{8cm}{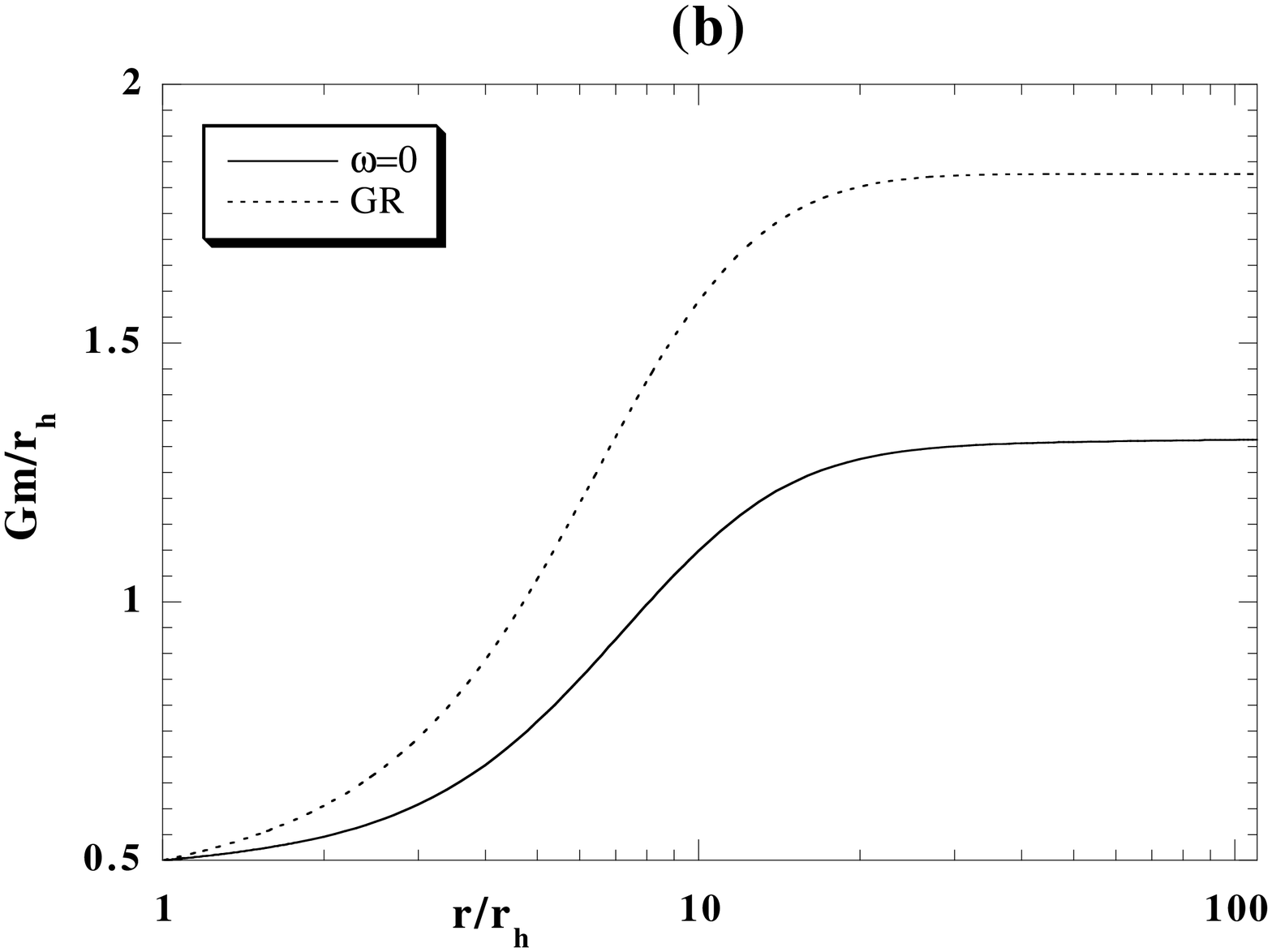}{}\\
\segmentfig{8cm}{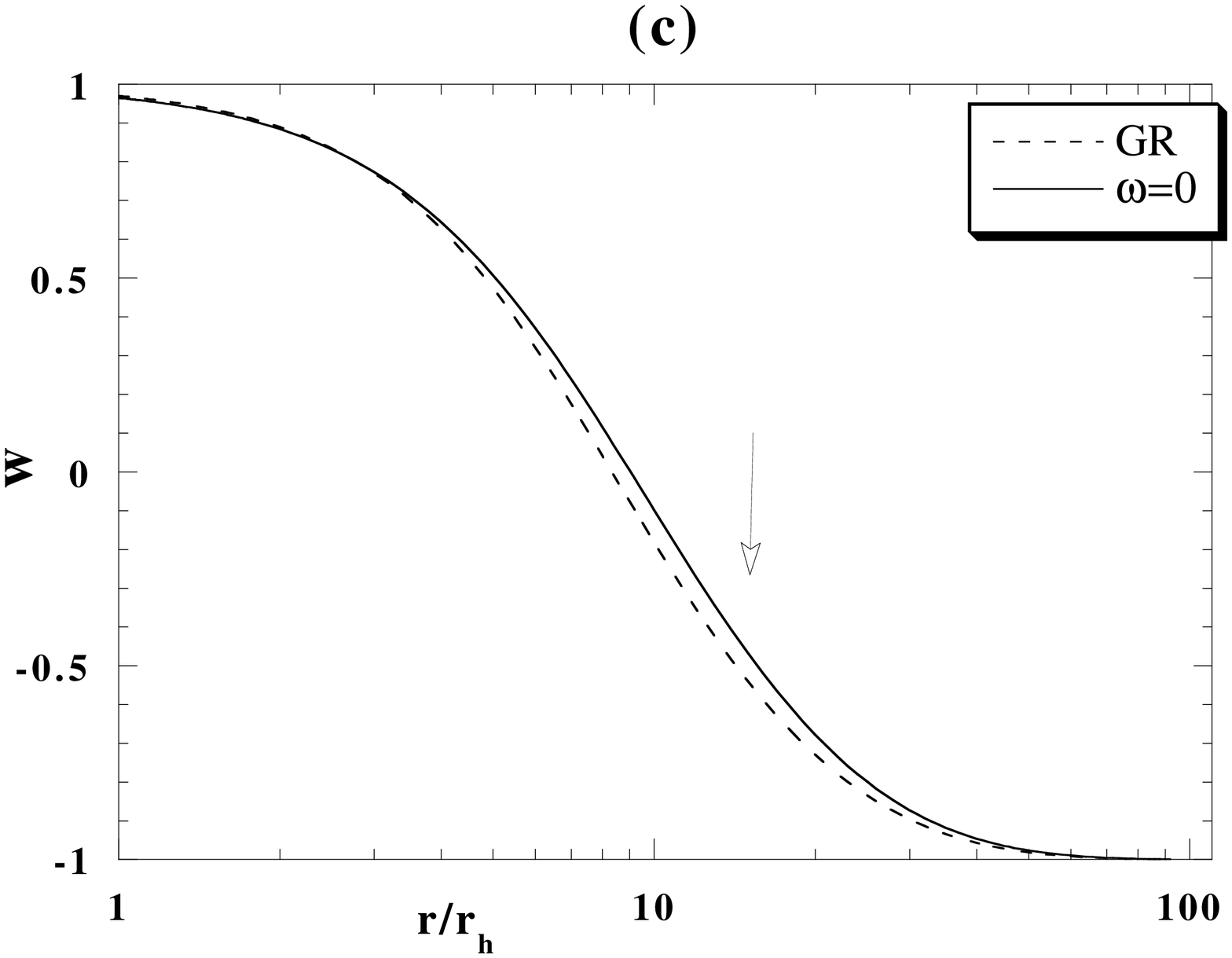}{}  
\segmentfig{8cm}{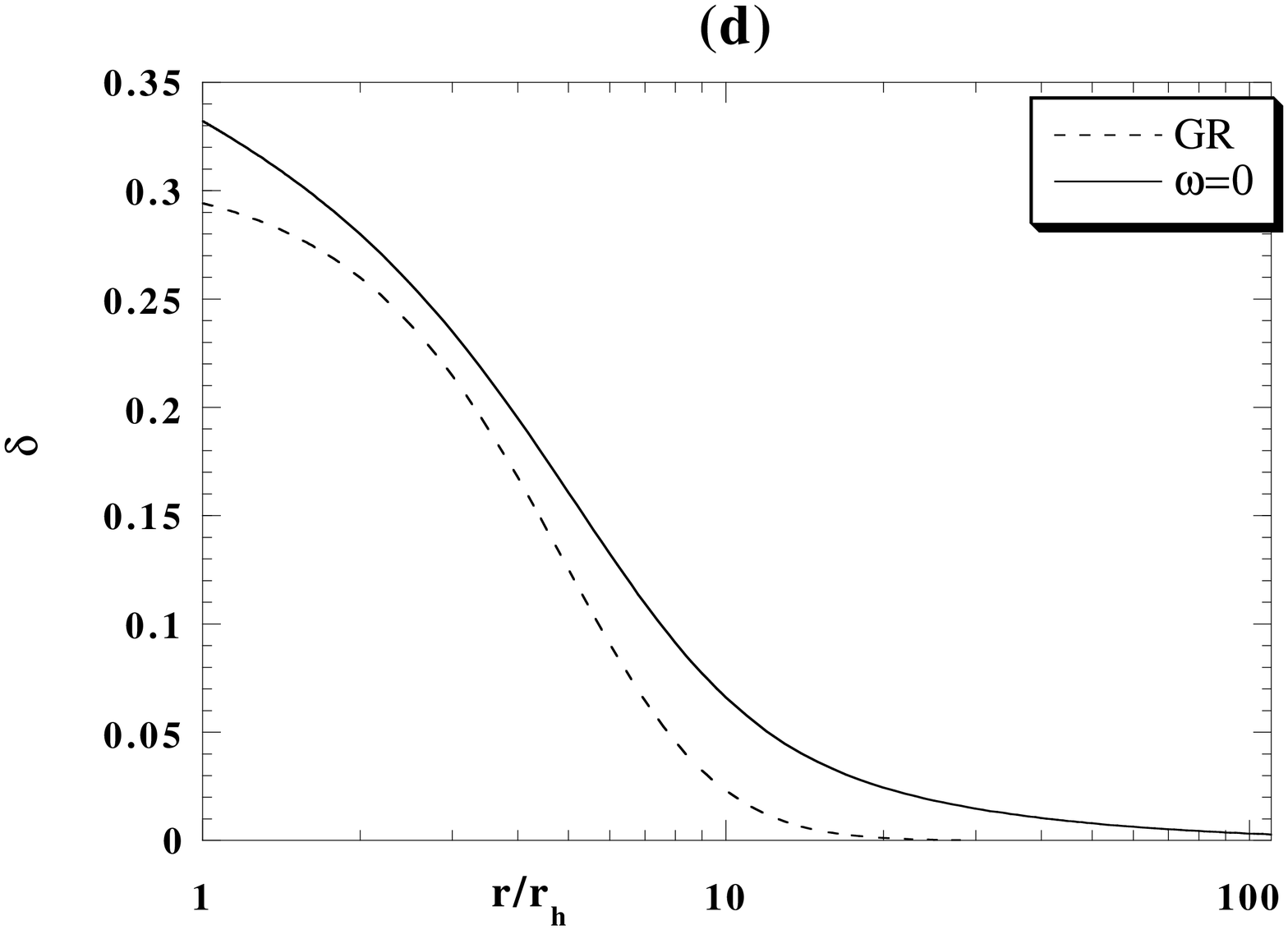}{}
\end{tabular}
\caption{ The solution of the Proca black hole in BD
theory with $\omega =0$ for $r_{h} = 0.5 l_{p}/g_{c}$ and $\mu = 0.15 g_{c} m_{p}$( (a) $\phi(r)$, (b) $\bar{m}(r)$, (c) w(r), (d) $\delta$(r) ). :  The Proca black hole in GR is also depicted as a reference by a dotted line.  The arrow in (c) shows
the Compton wavelength of the Proca field ($1/\mu$).  \label{solution} }
\end{figure}

As seen from Fig.1(a), the BD scalar field
decreases monotonically as
\begin{eqnarray}
\phi&\sim& \phi_{0}\left(1+\frac{2GM_s}{r}\right)\ ,
\label{eq:24}
\end{eqnarray}
where $M_s$ is  a constant and
called the scalar mass\cite{Will}.   Fig.1(c) shows that the
non-trivial structure of the non-Abelian field extends to the scale
of the Compton wavelength of the Proca field ($\sim 1/\mu$), which is
shown by an arrow.  From Fig.1(d), you may not see a clear
difference between the lapse function $\delta$ in BD theory and
that in GR, but $\delta$ in BD theory falls as $1/r$ as $r \rightarrow \infty$
while
$\delta$ in GR vanishes much faster  than
$1/r$.
In fact, from Eq.  (\ref{eq:8}) we find
\begin{eqnarray}
\frac{d\delta}{dr}\sim \frac{1}{\phi}\frac{d\phi}{dr}\
\sim -\frac{2GM_s}{r^2}
\label{eq:8zenkin}
\end{eqnarray}
near  spatial
infinity.
 This gives the relation between $M$ and $M_{g}$ as
\begin{eqnarray}
M_{g}=M+2M_s\ .  \label{eq:m-mg}
\end{eqnarray}

To see a property of a family of black hole solutions, we
show the relation between the gravitational mass $M_{g}$ and horizon
radius $r_h$ in   Fig.2.
The dotted lines denote the Proca black hole in GR with the same
parameters, i.e.
$\mu = 0.1 g_c m_p $ or $\mu = 0.15 g_c m_p$, and the
dot-dashed lines represent the Schwarzschild and colored black
holes, respectively, which we show as references.

As we mentioned in Fig.1(c), the non-trivial structure of the 
non-Abelian field is as large as  the scale of the Compton
wavelength   ($\sim 1/\mu$).  This is responsible for the existence
of a maximum horizon radius ($\sim 1/\mu$) as in GR.
 That is, beyond this critical horizon radius, a non-trivial
structure is swallowed into the horizon and then cannot
exist,  resulting in a trivial Schwarzschild spacetime.

The Schwarzschild black hole is a trivial solution ($m=M_{g},\
\delta=0,\ \phi=\phi_{0}
\ $and$\ w=-1$), which has no upper bound for a mass or a horizon
radius.  If the YM field is massless, a family of colored black
holes also exists as a non-trivial  black hole, where the  BD
scalar field is $\phi=\phi_{0}=$ constant. There is also no upper bound for horizon radius as in GR.

\begin{figure}
\begin{center}
\singlefig{12cm}{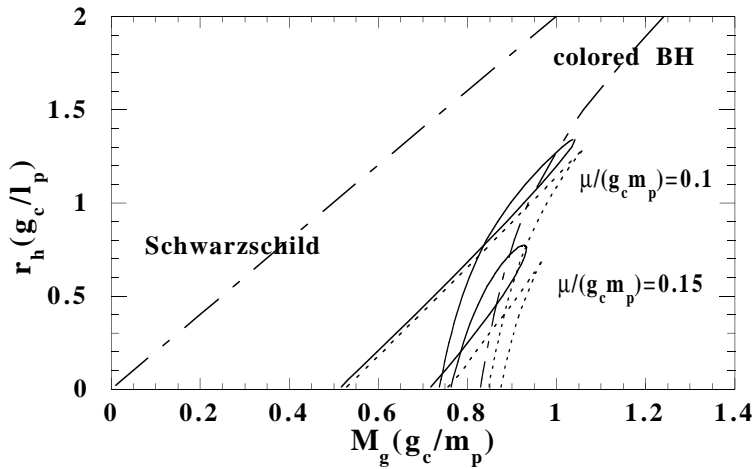}
\caption{$M_{g}-r_{h}$ diagram of the Proca black holes.
The solid loop lines denote Proca black holes with
$\mu=0.1 g_c m_p$ and $0.15 g_c m_p$ in BD theory ($\omega=0$).
  We  depict those   in  GR with
the same parameters by dotted lines.
The Schwarzschild and colored black holes are also shown as
references. \label{r-MBD}}
\end{center}
\end{figure}

The mass of the Proca black hole in BD theory is always
smaller than that in GR (see also Fig.1(b)). This is just because the value of the BD scalar field near the black hole is larger
than that at infinity, which means the effective gravitational
constant is always smaller than $G$.  Therefore the mass concentration
by gravitational attractive force  may get smaller.

In GR, there are two branches of black hole solutions: One is stable and the other is unstable.  Those two branches coincide at a
critical horizon radius or at a critical mass, where we find a cusp
structure on the gravitational mass $M_{g}$- horizon radius $r_{h}$ diagram. This cusp structure is a typical symptom of stability
change in catastrophe theory\cite{rf:CAT}.  The stability analysis by catastrophe theory agrees with that by linear perturbations.
\begin{figure}
\begin{center}
\singlefig{8cm}{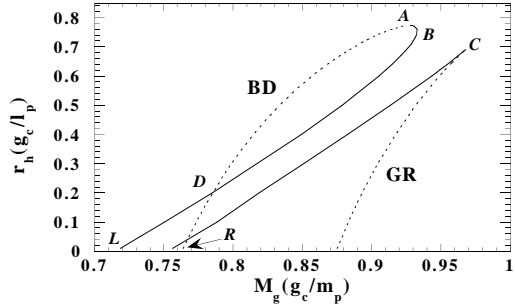}
\caption{$M_{g}$-$r_{h}$ diagram  for Proca black holes in BD
theory ($\omega=0$) and in GR. The mass of the Proca field is
$\mu=0.15g_cm_p$. \label{m-rm0.15}
}
\end{center}
\end{figure}
To see the detail and compare our solution with that in GR, we
depict  the enlarged diagram for  $\mu=0.15 g_c m_p$ in 
Fig.\ref{m-rm0.15}.
No cusp structure appears in BD theory. Though the solution curves seems to merge at the point D, we find two
solutions exist at D, which can be  distinguished  from field
distributions.   In GR, the maximum  points of horizon radius $r_{h}$ and of gravitational mass $M_{g}$ are the same, i.e., the point $C$ in
Fig.\ref{m-rm0.15}.  In BD theory, however
those two points,  $A$ (maximum horizon radius) and
$B$ (maximum gravitational mass), are different.  This result does
not depend on the choice of the  BD parameter $\omega$ and the mass of 
the Proca field $\mu$. In particular, a cusp structure  disappears in  BD
theory as mentioned above. One may wonder whether catastrophe
theory can  be simply applied to stability analysis as it is in BD theory, although  it works quite well in GR\cite{rf:Tachi}.
For fixing  $r_{h}$, we still have two  solutions in BD theory as in GR. Is there any
correspondence of those two solutions to two branches in GR?
We  expect that there are two types of black holes in the BD
theory as well.

As we discussed in our previous papers\cite{rf:Tachi}, if we divide the total energy density $\rho_{total}$ into  a kinetic term $\rho_{F^{2}}$ and a mass term $\rho_{A^{2}}$, one of the main differences between the two branches in GR (the solid-line and the dotted-line branches\footnote{In
\cite{rf:Tachi}, we  called them high-entropy and low-entropy
branches, respectively.}
in Fig.\ref{m-rm0.15}) comes from the difference of dominant
ingredient,  i.e., in the solid-line branch, $\rho_{A^{2}}$ is
dominant compared to $\rho_{F^{2}}$, stabilizing a black hole
solution. In the dotted-line  branch, the situation is opposite.  The
stable solid-line branch is Schwarzschild type, while the unstable
dotted-line branch is colored black hole type, in which the
non-Abelian field and gravity balance each other.

\begin{figure}[htbp]
\begin{tabular}{ll}
\segmentfig{8cm}{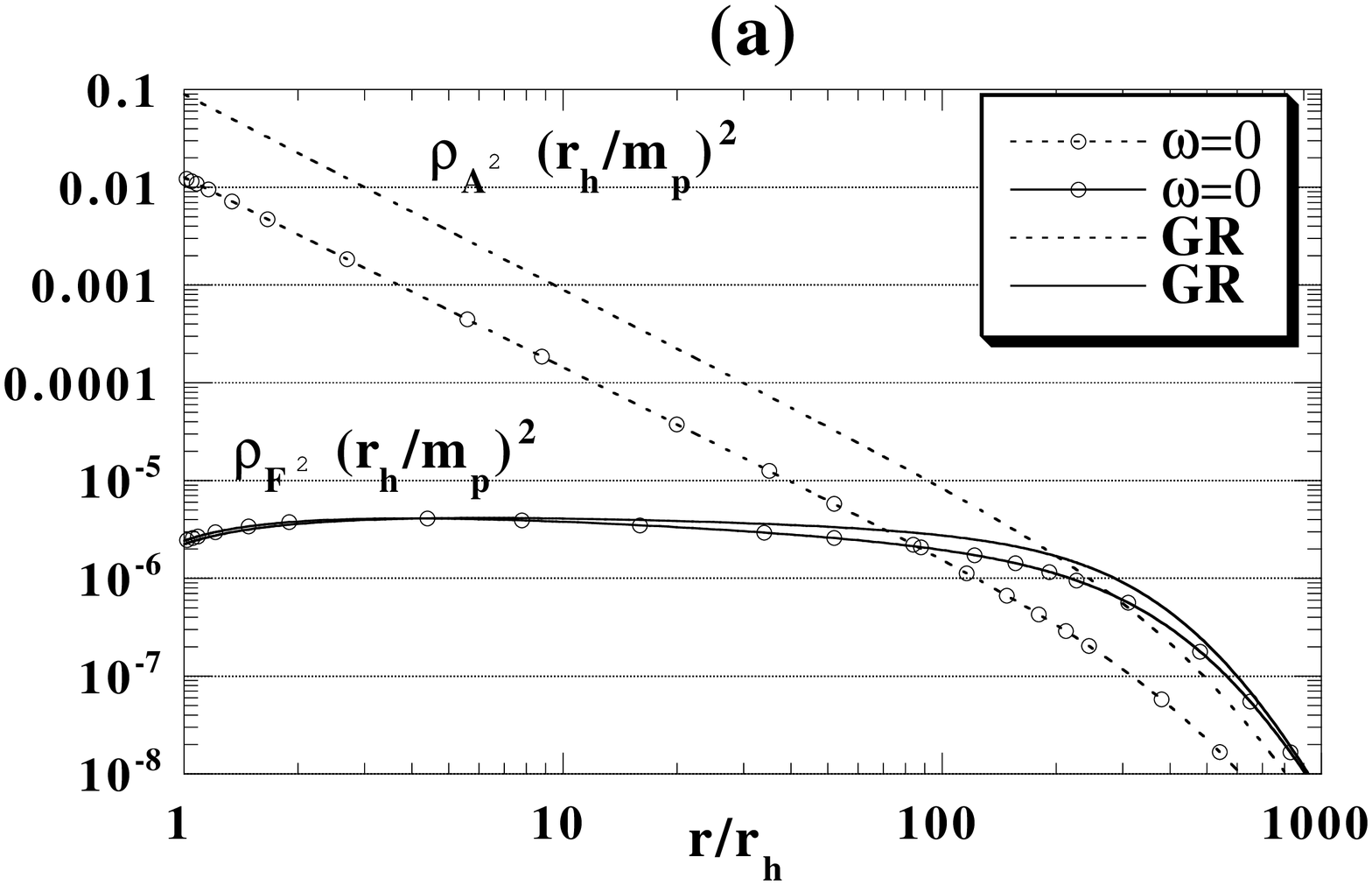}{}
\segmentfig{8cm}{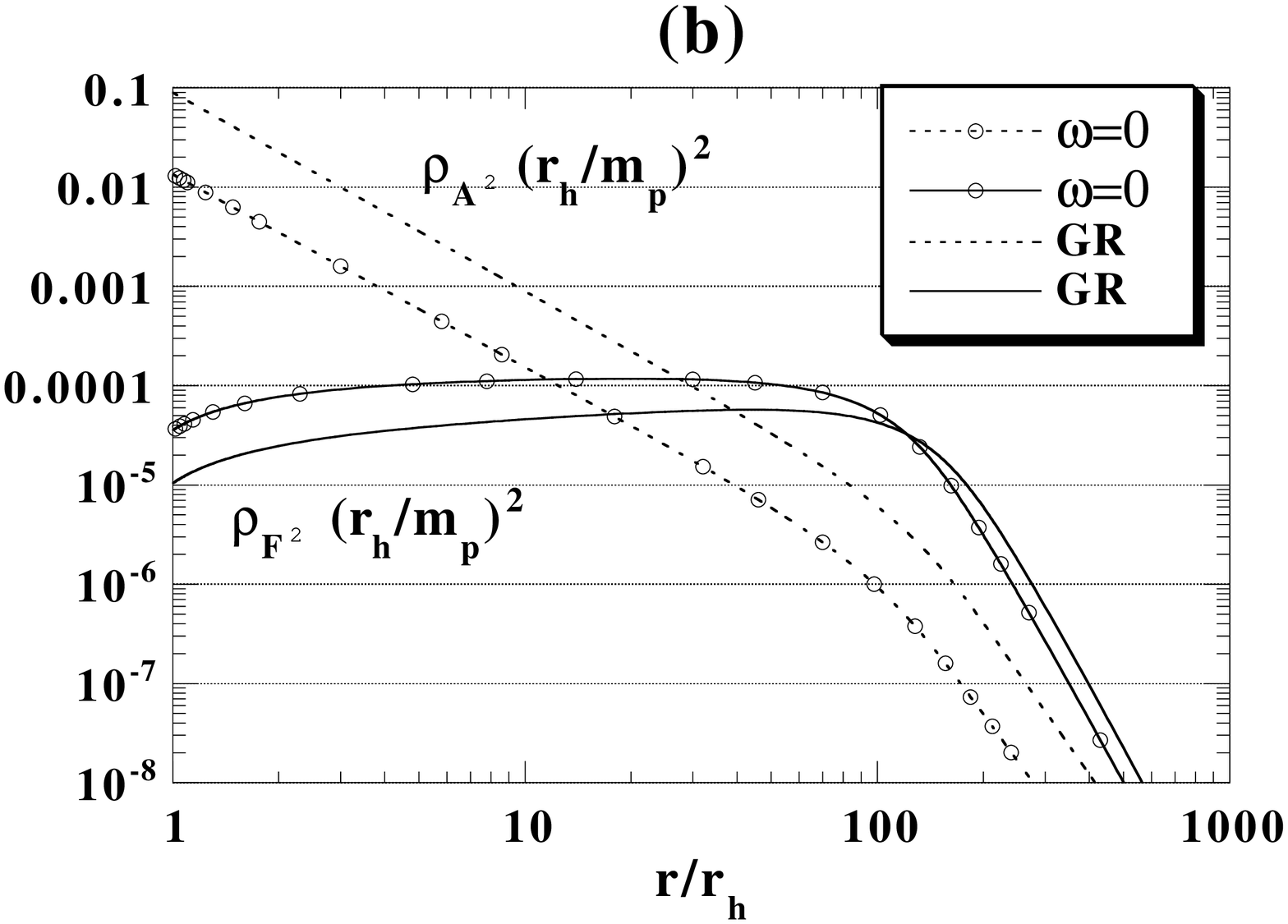}{}  
\end{tabular}
\caption{Distributions of the energy density for (a) the solid-line 
 and (b) the dotted-line branches of the Proca black holes with
$\mu=0.15g_cm_p$  both in  BD theory ($\omega=0$) and in GR.  The horizon radii of the black holes are $r_{h}=0.01 l_{p}/g_{c}$.     \label{energy}}
\end{figure}

In BD theory, if we divide the total energy density $\rho_{total}$ as 
\begin{eqnarray}
\rho_{total}&=&-T^{0}_{\ 0}=\rho_{A^{2}}+\rho_{F^{2}}+\rho_{\phi}\ ,
\label{eq:division}  
\end{eqnarray}
where  
\begin{eqnarray}
\rho_{A^{2}}\ ( \frac{r_{h} }{m_{p} } )^{2}&=&\frac{2(1+w)^{2}\bar{\mu}^{2}}{\bar{r}^{2}}
\left(2\phi+\bar{r}\frac{d\phi}{d\bar{r}} \right)^{-1}
\left\{
\frac{2\omega+1}{2\omega+3}
+\frac{2(\omega+1)\bar{r}}{(2\omega+3)\phi}
\frac{d\phi}{d\bar{r}}
\right\}\ ,
\label{eq:mass term}  \\
\rho_{F^{2}}\ ( \frac{r_{h} }{m_{p} } )^{2}&=&\frac{1}{\bar{r}^{2}\lambda_{h}^{2}}
\left(2\phi+\bar{r}\frac{d\phi}{d\bar{r}} \right)^{-1}
\left\{
2\left(1-\frac{2\bar{m}}{\bar{r}}\right)
\left(\frac{dw}{d\bar{r}}\right)^{2}
+\left( \frac{1-w^{2}}{\bar{r}}\right)^{2}
 \left(1+\frac{\bar{r}}{\phi}\frac{d\phi}{d\bar{r}} \right)
\right\}\ ,
\label{eq:kinetic term}  \\
\rho_{\phi}\ ( \frac{r_{h} }{m_{p} } )^{2}&=&\frac{1}{\bar{r}^{2}}
\left(2\phi+\bar{r}\frac{d\phi}{d\bar{r}} \right)^{-1}
\left\{ 
\frac{\omega+2}{2\phi}
\left(1-\frac{2\bar{m}}{\bar{r}}\right)
 \left(\bar{r}\frac{d\phi}{d\bar{r}}\right)^{2} 
-\bar{m}\frac{d\phi}{d\bar{r}}  
\right\}\ .         \label{eq:phi term}   
\end{eqnarray}  
We find the similar behavior to the case in GR, i.e., $\rho_{A^{2}}$
is dominant to $\rho_{F^{2}}$ in the solid-line
branch, while the opposite  is true in the dotted-line branch 
(see Fig.\ref{energy}).

\vspace{2cm}

In the solid-line branch, the black hole and non-Abelian structure are
rather independent. In fact, a particle-like solution in this
branch  can exist without gravity.  On the other hand, in the
dotted-line branch, we need both the non-Abelian field and  gravity.
Then,  we can divide the family of solutions into two:  a sold-line
branch from   $A$ to $L$
  (solid line) and a dotted-line branch  from  $R$
to $A$ (dotted line), respectively (see Fig.\ref{m-rm0.15}). In the
solid-line branch, the existence of the  BD scalar field  may not
change the black hole  structure, but it may affect a lot in the
dotted-line branch. This is because the non-Abelian field in the
solid-line branch does not give a dominant contribution to the 
black hole structure.   As we will see later, this becomes more
clear in the Einstein conformal frame,  in which the  effect of
the BD scalar field is reduced to matter coupling.

\begin{figure}
\begin{center}
\singlefig{8cm}{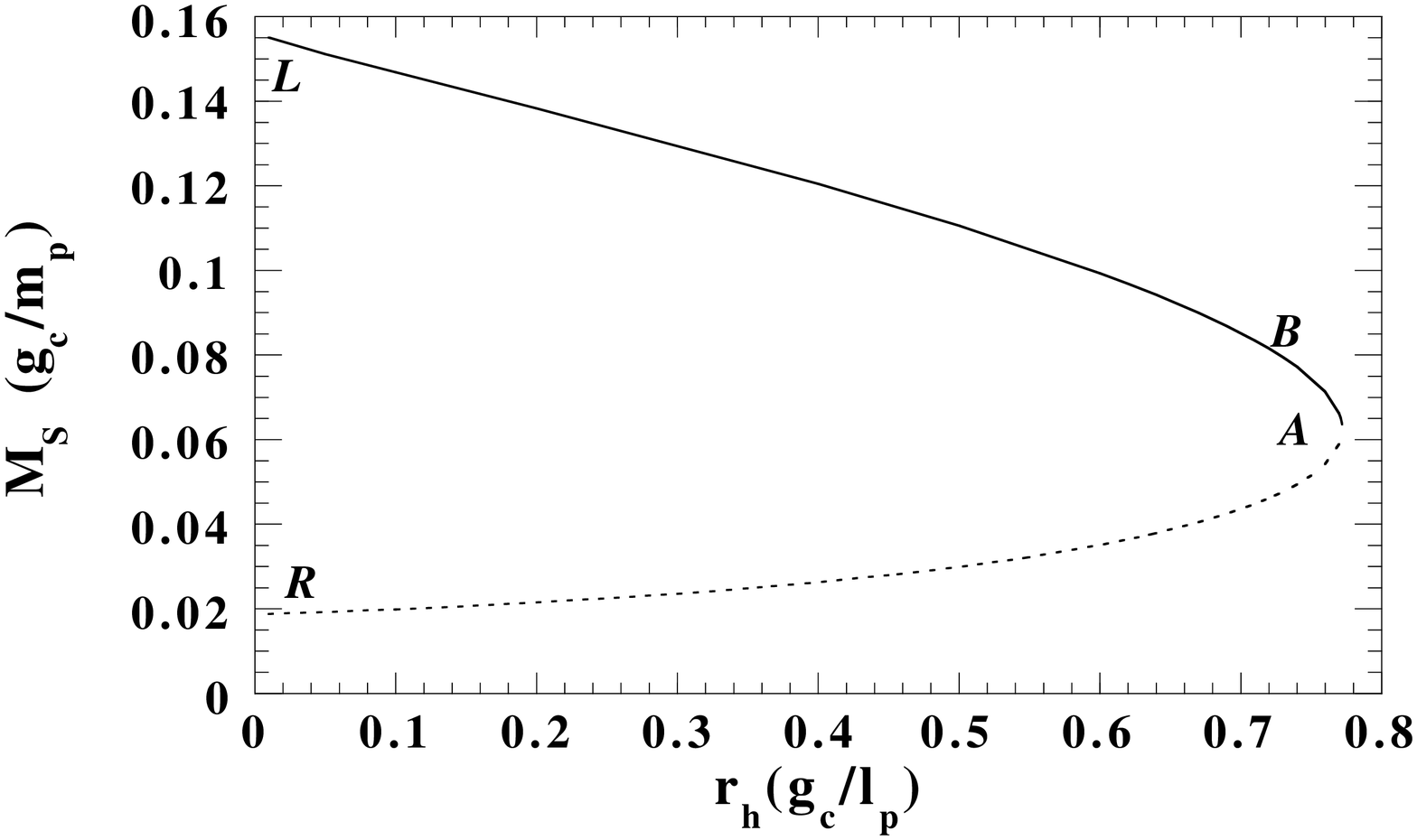}
\caption{$r_{h}$-$M_s$ (scalar mass) diagram of the Proca black holes with $\mu=0.15g_{c}m_{p}$  in BD theory ($\omega=0$).  The dotted  and
solid lines correspond to those in Fig.\protect\ref{m-rm0.15}.  \label{PS}
}
\end{center}
\end{figure}

\begin{figure}[htbp]
\begin{tabular}{ll}
\segmentfig{8cm}{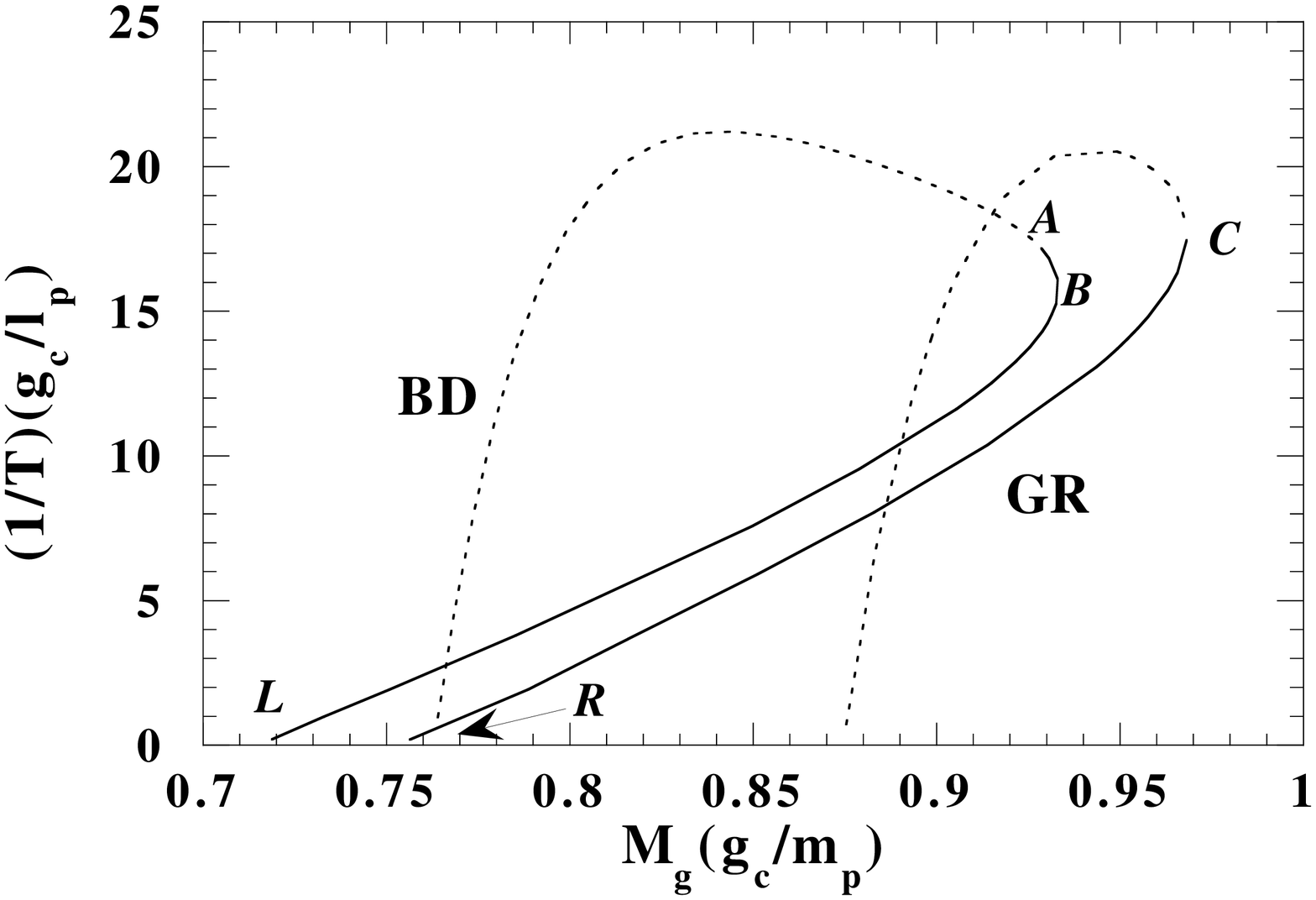}{(a)}
\hspace{5mm}
\segmentfig{8cm}{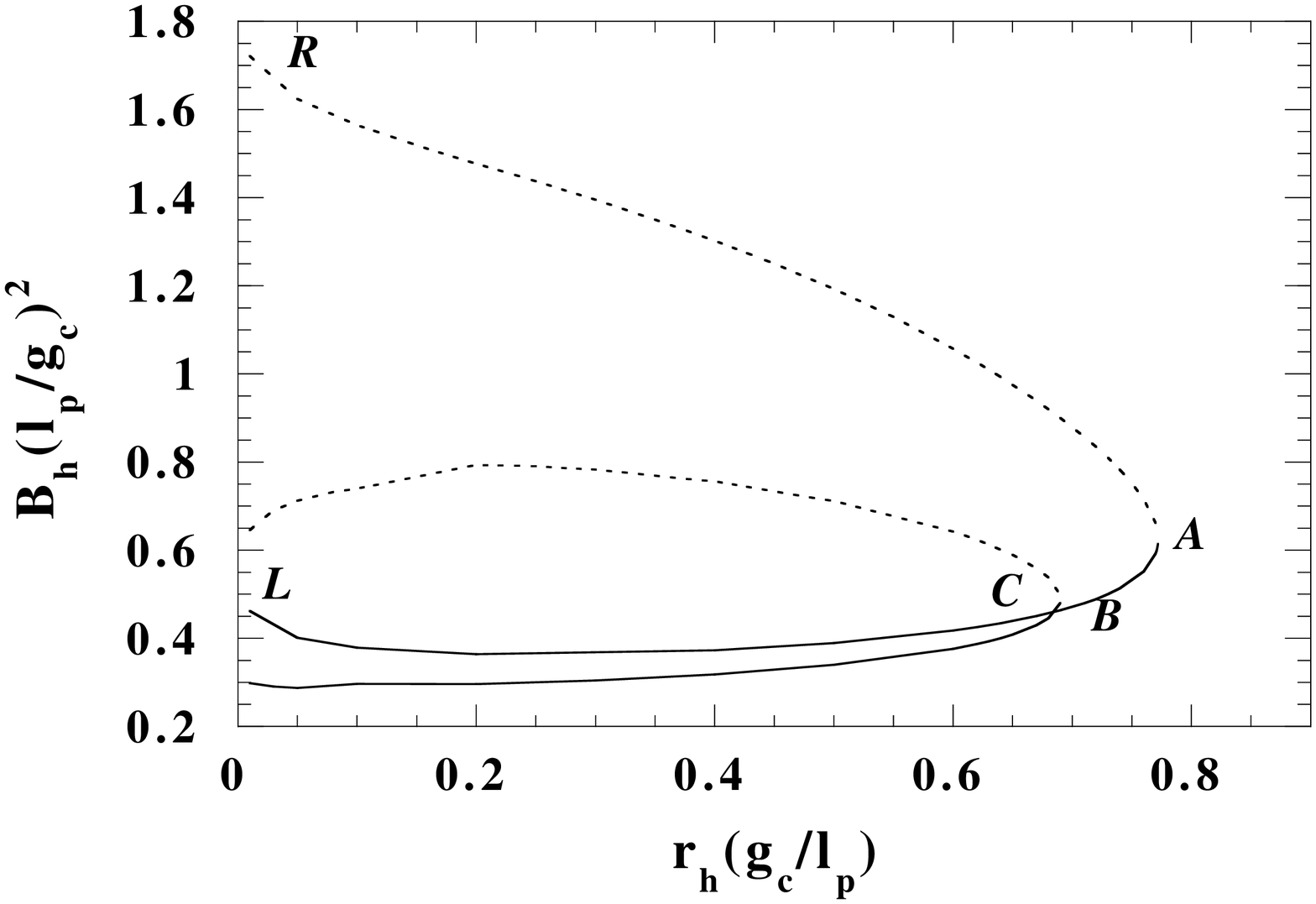}{(b)}
\end{tabular}
\vspace{5mm}
\caption{(a) $M_{g}$-$1/T$ diagram for Proca black holes with $\mu=0.15g_cm_p$ in BD theory ($\omega=0$) and in  GR.        \label{m-1Tm0.15}  }
\vspace{5mm}
\caption{(b) $r_{h}$-$B_{h}$ diagram for Proca black holes with $\mu=0.15g_cm_p$ in BD theory ($\omega=0$) and in  GR.        
\label{rh-strength}  }
\end{figure}

For these two branches, we  depict the scalar mass in terms of the horizon radius in Fig.\ref{PS}. The scalar mass
$M_s$ in the solid-line branch is always larger than that in
dotted-line branch.  We also show the inverse temperature $1/T$
in terms of $M_{g}$ and the field strength at the horizon $B_{h}$ 
in terms of $r_{h}$ of Proca  black holes  with $\mu=0.15 g_cm_p$ in Figs.\ref{m-1Tm0.15},\ref{rh-strength}, which are quite similar to those in GR. $T$ and $B_{h}$ are defined by 
\begin{eqnarray}
T & = & \frac{1}{4\pi r } e^{-\delta}(1-2G\frac{dm}{dr}) 
\bigg|_{\bar{r}=1}     \label{eq:temperature}  \\
B_{h} & = & ( {\rm Tr}\mbox{\boldmath$F$}^{2})^{1/2}
\bigg|_{\bar{r}=1}= \frac{ \sqrt{2}(1-w_{h}^{2}) }{r_{h}^{2}}  \ .      \label{eq:strength}  
\end{eqnarray}
Those also suggest that a stability may change somewhere in between
$A$ (maximum horizon radius) and 
$B$ (maximum gravitational mass).


\section{Proca Black Hole in the Einstein Conformal Frame}   %
The gravity part of BD theory is conformally equivalent  to
that of GR, and a description by use of  the Einstein conformal
frame sometimes gives us simpler basic equations and easier
analysis because the coupling of the BD scalar to gravity is moved
to a matter term and the gravity part is just described  as in the Einstein frame, which is already familiar.  Hence, here
we shall reanalyze our present problem in the Einstein conformal
frame. We consider a conformal transformation
\begin{eqnarray}
\hat{g}_{ab}=\frac{\phi}{\phi_{0}} g_{ab}\ .
\end{eqnarray}
The equivalent action $\hat{S}\equiv S/\phi_0 $ is given as
\begin{eqnarray}
\hat{S} & = & \displaystyle\int d^{4}x \sqrt{-\hat{g}}\left[
\frac{1}{2\kappa^{2}} \hat{R} -
\frac{1}{2}\hat{\nabla}_{\alpha}\varphi\hat{\nabla}^{\alpha}\varphi
-{1 \over \phi_{0}}\left(\frac{1}{16\pi g_{c}^{2}} {\rm
Tr}\mbox{\boldmath
$F$}^{2}+\frac{\mu^{2}}{8\pi g_{c}^{2}}\exp(-\kappa\beta\varphi)
{\rm Tr}\mbox{\boldmath
$A$}^{2}\right)\right]\ ,       \label{eq:20}  \\[.5em]
&&\ \ \ \
\varphi\equiv
\frac{1}{\kappa\beta}\ln \left({\phi \over
\phi_{0} }\right) \ , \ \ \ \ \ \ \ \ \beta\equiv\left(
\frac{2\omega+3}{2}
\right)^{-1/2}.
\label{eq:21}
\end{eqnarray}
 For a black hole  solution, if we define  spherically symmetric
coordinates in the Einstein frame   as
\begin{eqnarray}
d\hat{s}^{2}&\equiv &\frac{\phi}{\phi_{0}}ds^{2}   \nonumber  \\
&=&-\left[1-\frac{2G\hat{m}(\hat{r})}{\hat{r}}\right]
e^{-2\hat{\delta} (\hat{r})}dt^{2}+
\left[1-\frac{2G
\hat{m}(\hat{r})}{\hat{r}}\right]^{-1}d\hat{r}^{2}
+\hat{r}^{2}d\Omega^{2}\ ,
\label{eq:2.1_E}
\end{eqnarray}
we find
\begin{eqnarray}
\hat{M}\equiv \lim_{\hat{r}\rightarrow \infty} \hat{m}(\hat{r}) =M_{g}-M_s,\ \ \
\hat{r}_{h}=r_{h}\sqrt{\frac{\phi_{h}}{\phi_{0}}}\ ,
\label{eq:18}
\end{eqnarray}
where variables with a caret  $\hat{}$  denote those in the
Einstein frame.  We also introduce dimensionless
variables and parameters as
\begin{eqnarray}
\bar{\hat{r}}=\hat{r}/\hat{r}_{h}\ ,\ \bar{\hat{m}}
=G\hat{m}/\hat{r}_{h}
\ , \ \hat{\lambda}_{h}=\hat{r}_{h}g_{c}/l_{p}\ ,\ \bar{\varphi}
=\varphi/m_{p}\ .
\label{eq:pkikaku1}
\end{eqnarray}
The basic equations are now
\begin{eqnarray}
\frac{d\bar{\hat{m}}}{d\bar{\hat{r}}}&=& \left\{
\frac{1}{\phi_{0}\hat{\lambda}_{h}^{2} }
\left( \frac{dw}{d\bar{\hat{r}}} \right)^{2}+2\pi
\bar{\hat{r}}^{2}
\left( \frac{d\bar{\varphi}}{d\bar{\hat{r}}} \right)^{2} \right\}
\left( 1-\frac{2\bar{\hat{m}}}{\bar{\hat{r}}} \right)
+\frac{1}{ 2\phi_{0}\hat{\lambda}_{h}^{2} }
\left( \frac{1-w^{2}}{\bar{\hat{r}}} \right)^{2}  \nonumber  \\
&&+\frac{\bar{\mu}^{2}}{\phi_{0}}\exp{
(-\sqrt{8\pi}\beta\bar{\varphi}) }(1+w)^{2}    \ ,
\label{eq:peqefra1}   \\
\frac{  d\hat{\delta}  }{  d\bar{\hat{r}}  }&=&
-4\pi\bar{\hat{r}}
\left( \frac{  d\bar{\varphi}  }{  d\bar{\hat{r}}  }  \right)^{2}
-\frac{2}{  \phi_{0}\hat{\lambda}_{h}^{2}\bar{\hat{r}}   }\left(
\frac{dw}{d\bar{\hat{r}}} \right)^{2}  \ ,
\label{eq:peqefra2}   \\
\frac{d^{2}\bar{\varphi}}{d\bar{\hat{r}}^{2}}& =& -\frac{2}{\bar{\hat{r}} } +\left( 1-\frac{  2\bar{\hat{m}}  }{\bar{\hat{r}}  }  \right)^{-1}
\left[
\left(  \frac{  d\bar{\varphi}  }{  d\bar{\hat{r}}  }  \right)
\left\{
-\frac{  2\bar{\hat{m}}  }{  \bar{\hat{r}}^{2}  }+
\frac{  \bar{\hat{r}}  }{  \phi_{0}\hat{\lambda}_{h}^{2}  }
\left( \frac{  1-w^{2}  }{  \bar{\hat{r}}^{2}  }  \right)^{2}
+2\bar{\hat{r}} \frac{  \bar{\mu}^{2}  }{  \phi_{0}  }
\exp{ (-\sqrt{8\pi}\beta\bar{\varphi}) }
\left( \frac{  1+w  }{  \bar{\hat{r}}  }  \right)^{2}
\right\}
\right.     \nonumber  \\
&&\left.
-\frac{ \bar{\mu}^{2}\beta\sqrt{8\pi} }{ 4\pi\phi_{0} }
\exp{ (-\sqrt{8\pi}\beta\bar{\varphi}) }\left(    \frac{1+w}{
\bar{\hat{r}} }
\right)^{2}
\right] \ ,
\label{eq:peqefra3}   \\
\frac{d^{2}w}{d\bar{\hat{r}}^{2}}&=& \left(
1-\frac{2\bar{\hat{m}}}{\bar{\hat{r}}}\right)^{-1}
\left[
\frac{dw}{d\bar{\hat{r}}}
\left\{
-\frac{2\bar{\hat{m}}}{\bar{\hat{r}}^{2}}
+\frac{\bar{\hat{r}}}{\phi_{0}\hat{\lambda}_{h}^{2}}
\left( \frac{1-w^{2}}{\bar{\hat{r}}^{2}}\right)^{2}+
2\bar{\hat{r}} \frac{\bar{\mu}^{2}}{\phi_{0}} \exp{
(-\sqrt{8\pi}\beta\bar{\varphi}) }
\left( \frac{1+w}{\bar{\hat{r}}}\right)^{2}
\right\}\right.    \nonumber  \\
&&\left. -\frac{w(1-w^{2})}{\bar{\hat{r}}^{2}}
+\bar{\mu}^{2}\hat{\lambda_{h}^{2}}
\exp{ (-\sqrt{8\pi}\beta\bar{\varphi}) }(1+w)
\right] \ .  \label{eq:peqefra4}
\end{eqnarray}
As we expected, these are
simpler than those described  in the BD frame.
The boundary conditions are similar  to the ones in the BD frame.
From our numerical calculation, we can show that
$\hat{M}=\hat{M}_{g}$ because $\hat{\delta}$ vanishes faster than
$\hat{r}^{-1}$.

\begin{figure}[htbp]
\begin{tabular}{ll}
\segmentfig{8cm}{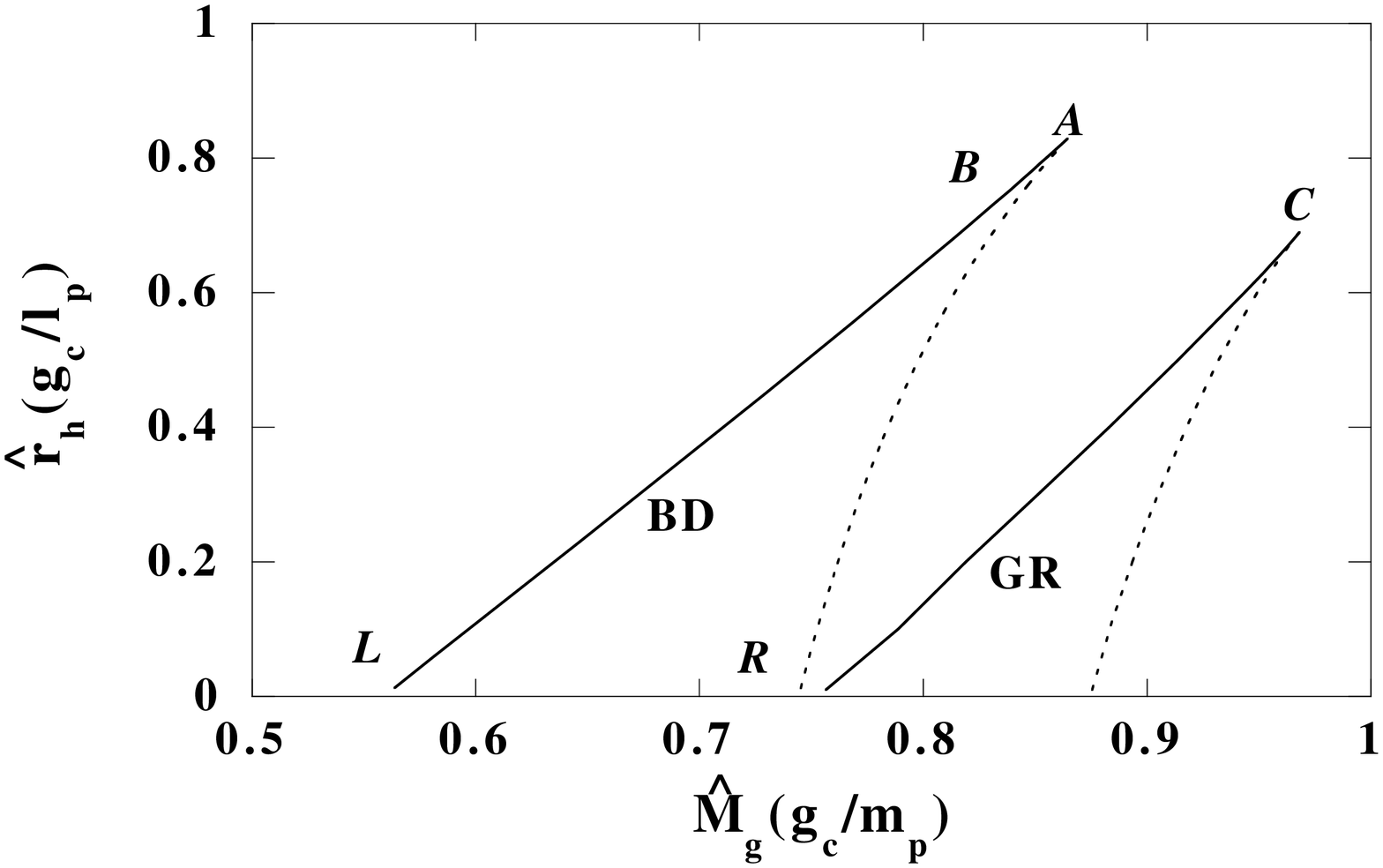}{(a)}
\hspace{5mm}
\segmentfig{8cm}{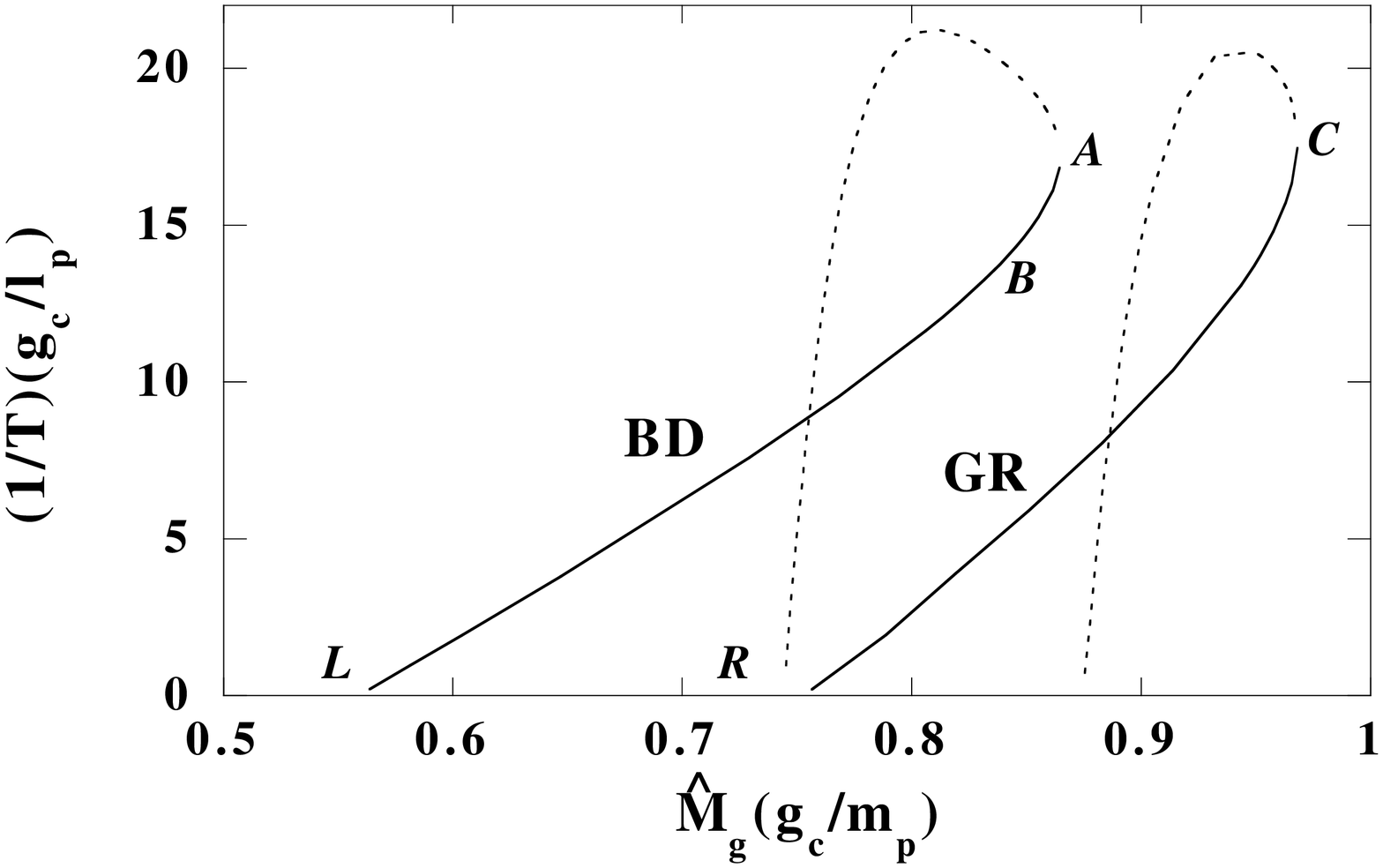}{(b)}
\end{tabular}
\vspace{5mm}
\caption{(a) $\hat{M}_{g}$-$\hat{r}_{h}$ diagram in the Einstein frame 
for Proca black holes in BD theory ($\omega=0$) and in GR. 
The mass of the Proca field is $\mu=0.15g_{c}m_{p}$.
We find a cusp structure, which indicates a stability
change via catastrophe theory.
\label{7}}
\vspace{5mm}
\caption{(b) $\hat{M}_{g}$-$1/T$ diagram in the Einstein frame 
for Proca black holes in BD theory ($\omega=0$) and in GR. 
The mass of the Proca field is $\mu=0.15g_{c}m_{p}$.
\label{8}   }
\end{figure}

First, in Fig.\ref{7}, we show the 
$\hat{M}_{g}$-$\hat{r}_{h}$ diagram in the Einstein frame that is 
related to Fig.\ref{m-rm0.15} by conformal transformation.
Surprisingly, we recover a cusp structure even in BD theory.  The solid-line is always located above the
dotted-line branch as in GR.  We also show the inverse
temperature
$1/T$ in terms of the gravitational mass $\hat{M}_{g}$ in
Fig.\ref{8}. Both figures show that the properties of the Proca black holes in BD theory are quite similar to those in GR.
\vspace{1cm}
This suggests that catastrophe theory will be  simply applied in a stability analysis for non-Abelian black holes  in  BD theory as well.

 From the point of view of catastrophe theory\cite{rf:CAT}, stability
changes at a cusp point in the control parameter-potential
function diagram. In GR,  if we regard  gravitational mass and 
black hole entropy (or equivalently the area of the event horizon) as a
control parameter and a potential function, respectively, we find a
cusp $C$ in the $M_g$-$r_h$ ($\hat{M}_g$-$\hat{r}_h$) diagram
(Figs.\ref{m-rm0.15},\ref{7}), which is a symptom of stability change in catastrophe theory.   In fact the stability of the black hole does change at this cusp point $C$. In BD theory, however, a cusp structure does
not  appear in  Fig.\ref{m-rm0.15},  while it does in Fig.\ref{7}.
This suggests that if we use the variables in the
Einstein frame, we can simply apply catastrophe theory to the stability analysis in BD theory as it is. From Fig.\ref{7}, catastrophe theory predicts  that stability change can occur at the point
$A$.  From Fig.\ref{m-rm0.15}, however, no such prediction is possible.

To study stability, we have another method, i.e., a turning point
method for  thermodynamical variables\cite{rf:Katz}. Stability
will change  at the point where $d(1/T)/dM=\infty$.
In GR, we understand that a stability change  occurs at the point $C$
in Figs.\ref{m-1Tm0.15},\ref{8}. This is consistent with analysis by catastrophe theory. In BD theory,  $d(1/T)/dM=\infty$ occurs at the point $B$ in Fig.\ref{m-1Tm0.15} (BD frame),
which is inconsistent with the stability analysis by catastrophe
theory. However, if we use the $\hat{M}_g$-$1/T$ diagram in the Einstein
frame, the divergence occurs at the point $A$, which is
consistent with catastrophe theory. To understand this
inconsistency, we have to remember that  variables in the
turning point method should satisfy thermodynamical laws, in
particular the ``mass" of a black hole should satisfy the first law
of  black hole thermodynamics. In fact,  the gravitational mass in  the BD frame does not satisfy the first law of black hole thermodynamics,
while it does so  for the variables in the Einstein frame
(Fig.\ref{8})\footnote{We can show that thermodynamical variables
in the Einstein frame satisfy the first law.\protect\cite{MTT}},
and therefore the turning point  method could be  applied.  We
expect  that a stability change occurs at the point $A$ but not
at the point $B$, and this is  consistent  with catastrophe  theory.

These conjectures for stability should be justified by analyzing 
linear perturbations of black holes and  black hole
thermodynamics\cite{MTT}.

\section{The Effects of the Brans-Dicke Scalar Field}             %
By  use of the Einstein frame, we also understand easily a
qualitative difference between black holes in BD theory and in
GR. As we see in the action (\ref{eq:20}), the coupling of the BD
scalar field appears in the mass term.  This coupling reduces
effectively the mass of the Proca field by a factor
$\exp(-\kappa\beta\varphi/2)$ because $\varphi$ is
monotonically decreasing to zero as $r \rightarrow \infty$ (see
Fig.\ref{solution}(a)).  In GR, as the mass is reduced, the Proca black hole
shifts in the left-upper direction in the $M_{g}$-$r_{h}$ diagram (see
Fig.\ref{r-MBD}).  In fact, in the  limit of zero mass, we recover
Schwarzschild and colored black hole branches.  As a result, for a
fixed  Proca field mass $\mu$, the black hole solution in BD theory also shifts in the left-upper direction from that in GR because of the coupling.
Another contribution is $\phi_{0}$, which appears in the matter
Lagrangian. Since $\phi_{0}^{-1}$ is its overall factor, this effect
is renormalized by a redefinition of the gauge coupling constant
$g_{c}$, i.e., $g_{c}'=\sqrt{\phi_{0} } g_{c}$.
As $\phi_{0}$ changes monotonically from 1 to $\infty$ for
$\infty>\omega>-3/2$, the effective gauge coupling
constant $g_{c}'$  changes from $g_{c}$ to $\infty$.

The effects of the BD scalar field are divided into two:\\
(1) The gauge coupling constant is renormalized as
$g_{c}'=\sqrt{\phi_{0}} g_{c}$, which gives a stronger coupling than
that in GR.\\
(2) The Brans-Dicke scalar field decreases as $r \rightarrow
\infty$, which gives an effective change of the mass of the Proca
field, i.e., $\mu '=\mu \exp(-\kappa\beta\varphi/2)$.\\

\begin{figure}[htbp]
\begin{tabular}{ll}
\segmentfig{8cm}{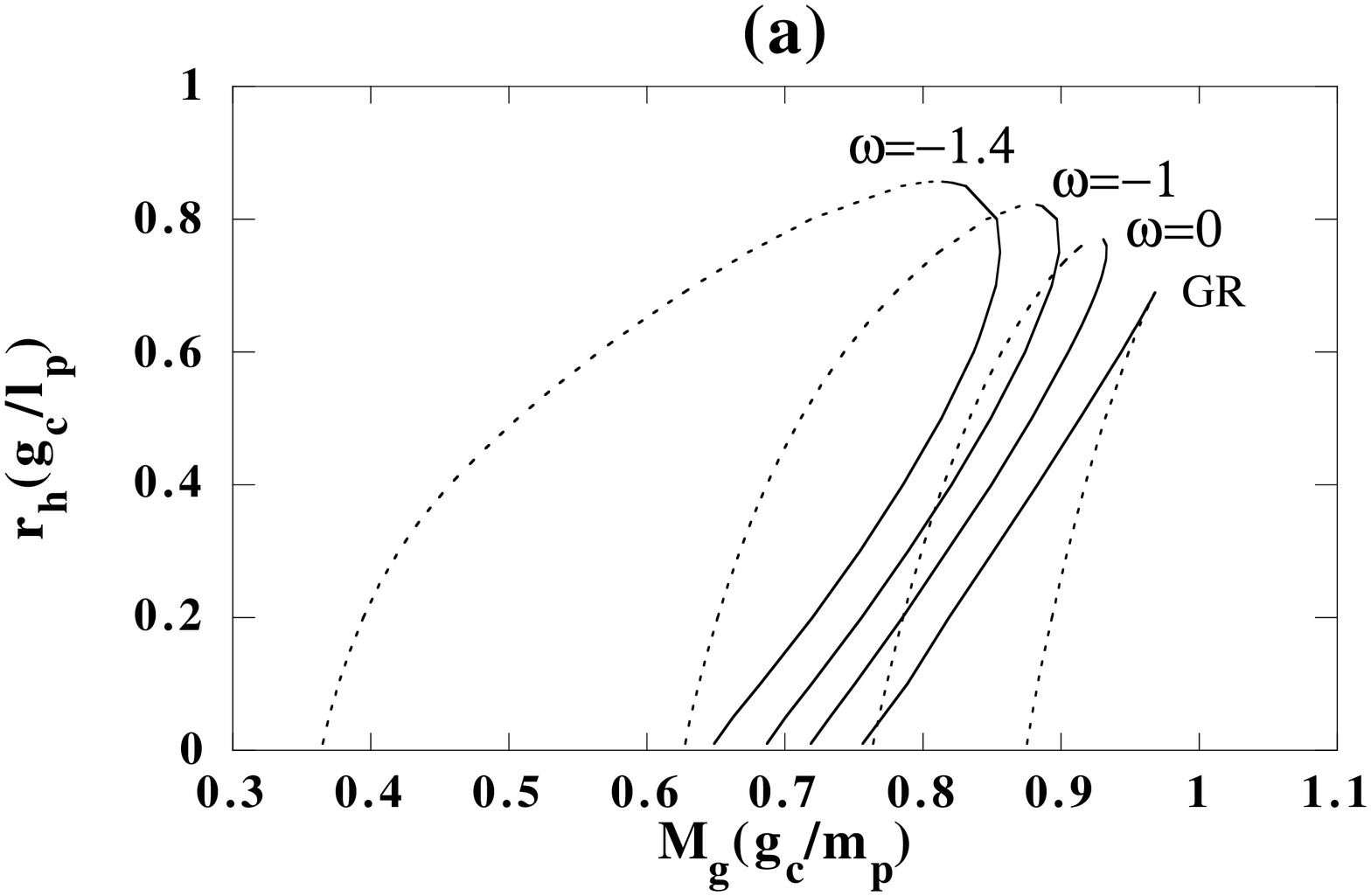}{}
\segmentfig{8cm}{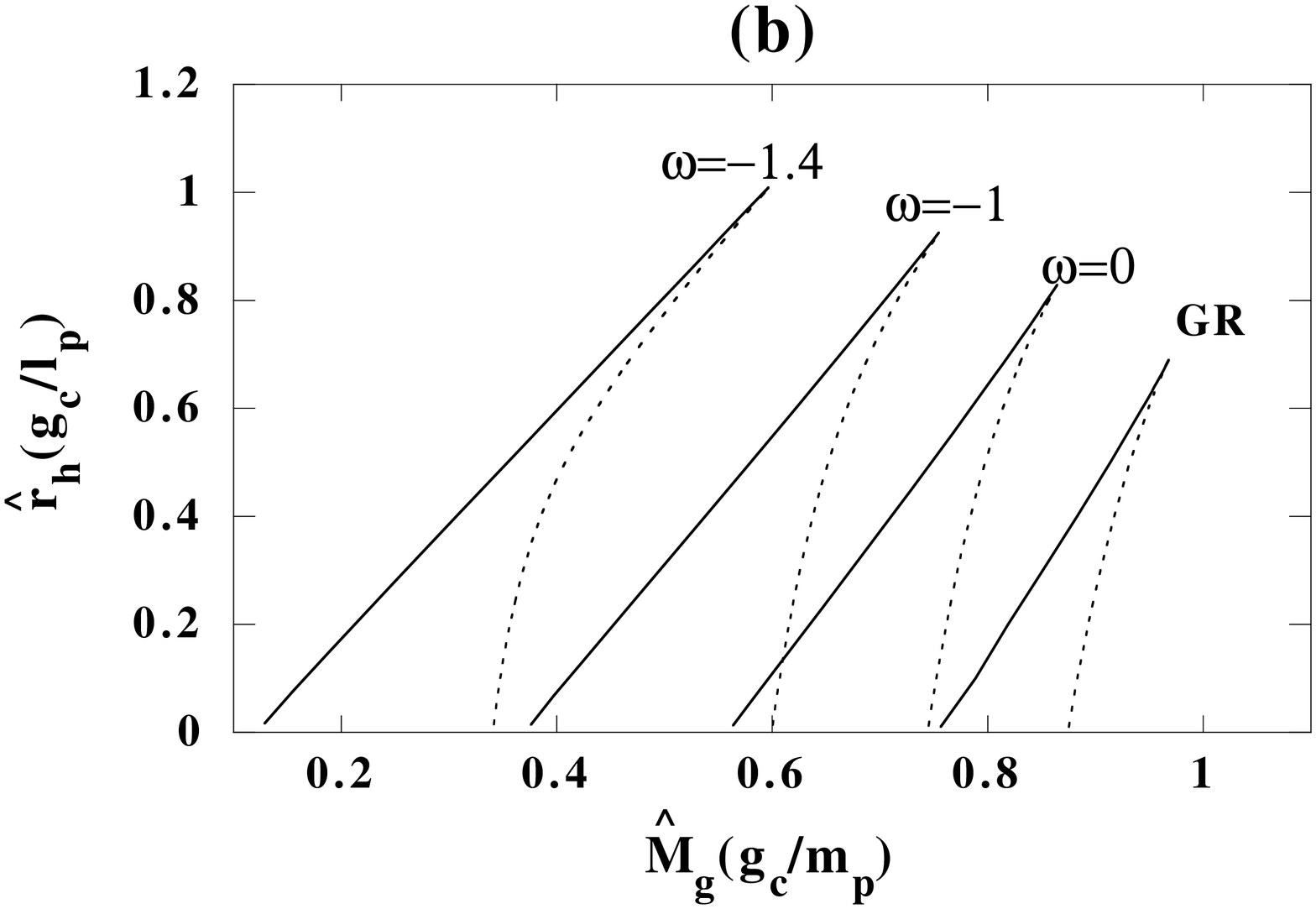}{}
\end{tabular}
\caption{
(a) $M_{g}$-$r_{h}$ diagram in the BD frame 
(b) $\hat{M}_{g}$-$\hat{r}_{h}$ diagram in the Einstein frame for
several values of $\omega$. The mass of the Proca field is 
$\mu=0.15g_{c}m_{p}$. \label{wdem-rP}
}
\end{figure}

\begin{figure}[htbp]
\begin{tabular}{ll}
\segmentfig{8cm}{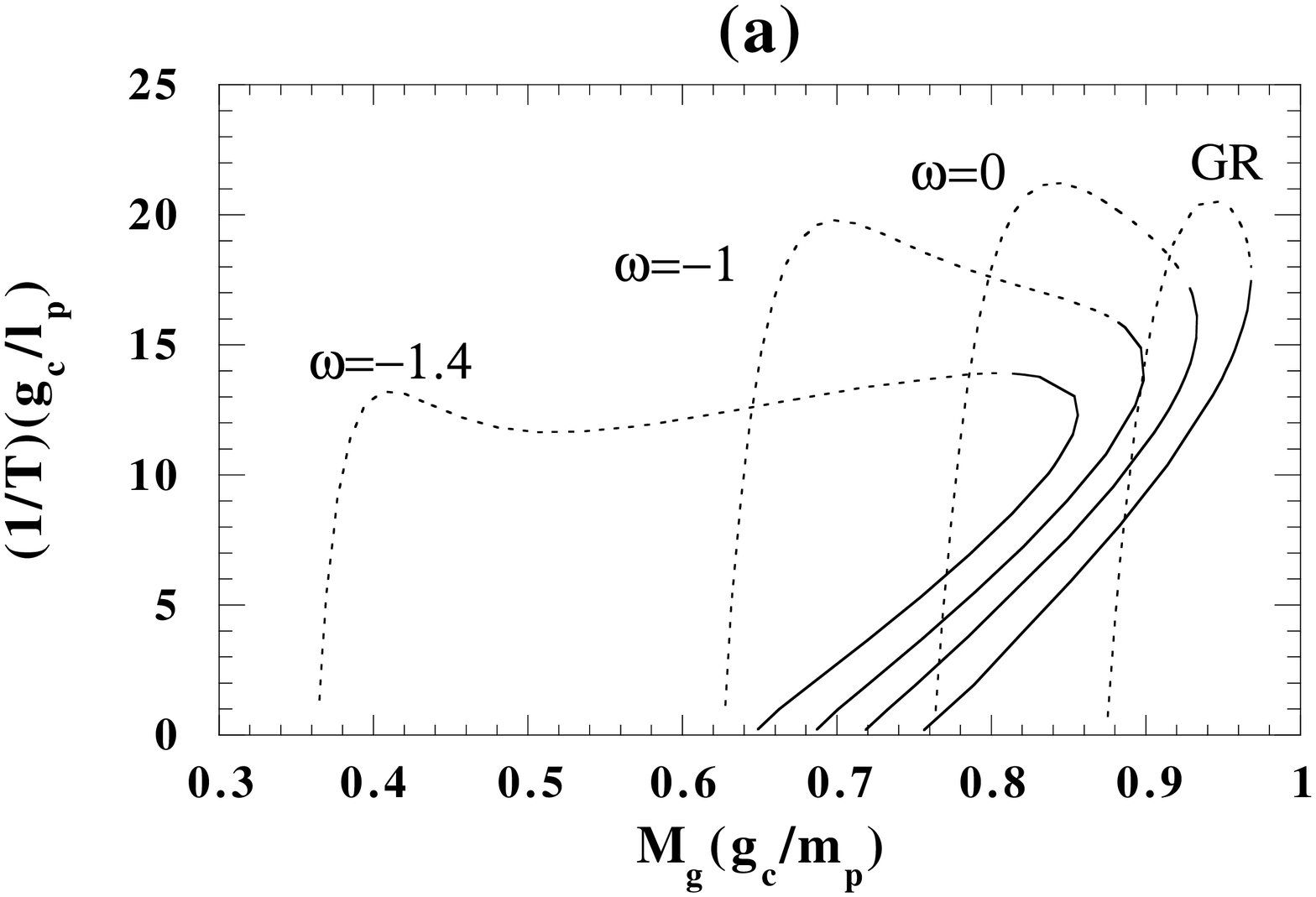}{}
\segmentfig{8cm}{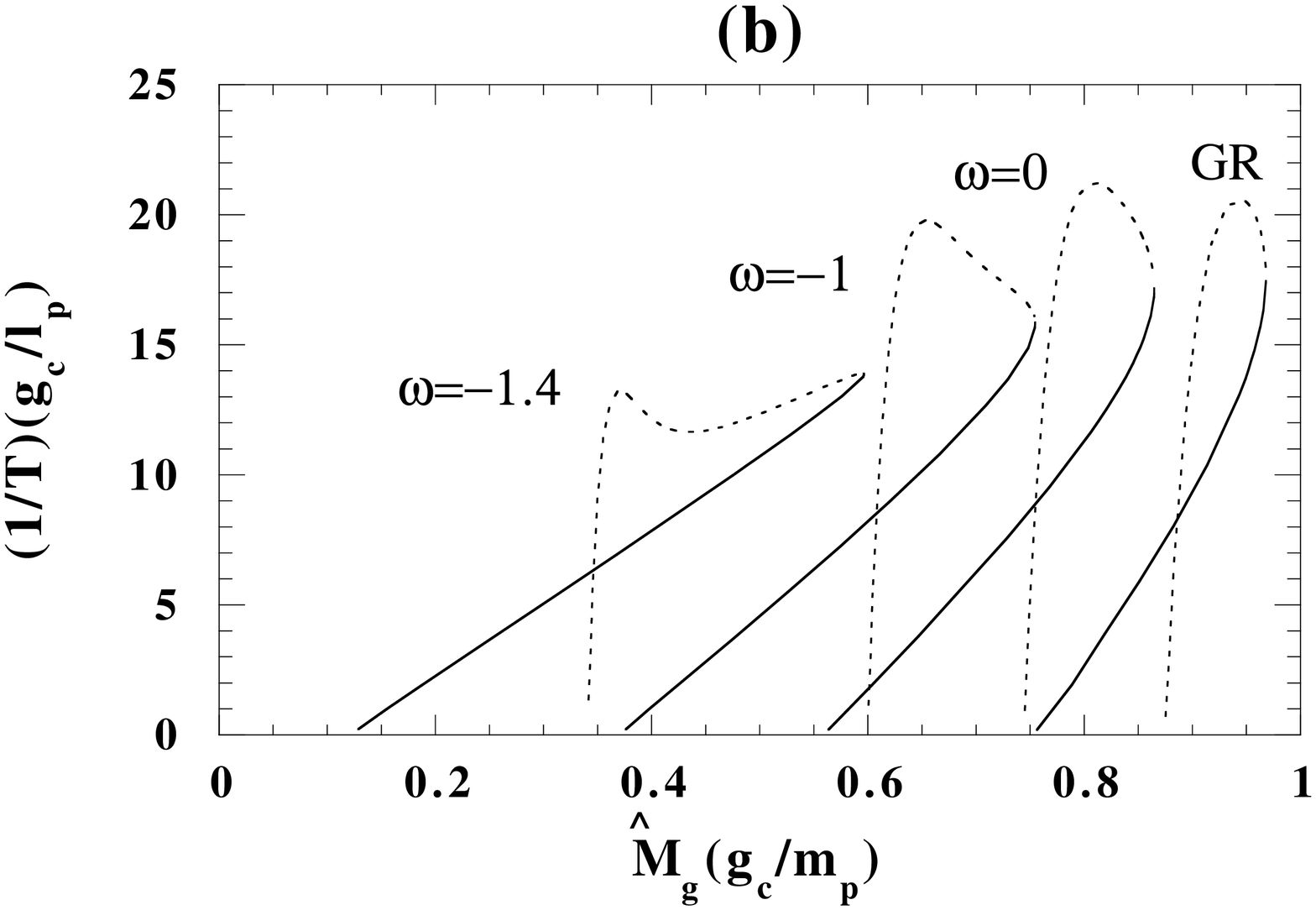}{}
\end{tabular}
\caption{
(a) $M_{g}$-$1/T$ diagram in the BD frame 
(b) $\hat{M}_{g}$-$1/T$ diagram in the Einstein frame for
several values of $\omega$. The mass of the Proca field is 
$\mu=0.15g_{c}m_{p}$. \label{wdem-1TP}
}
\end{figure}

\begin{figure}[htbp]
\begin{tabular}{ll}
\segmentfig{8cm}{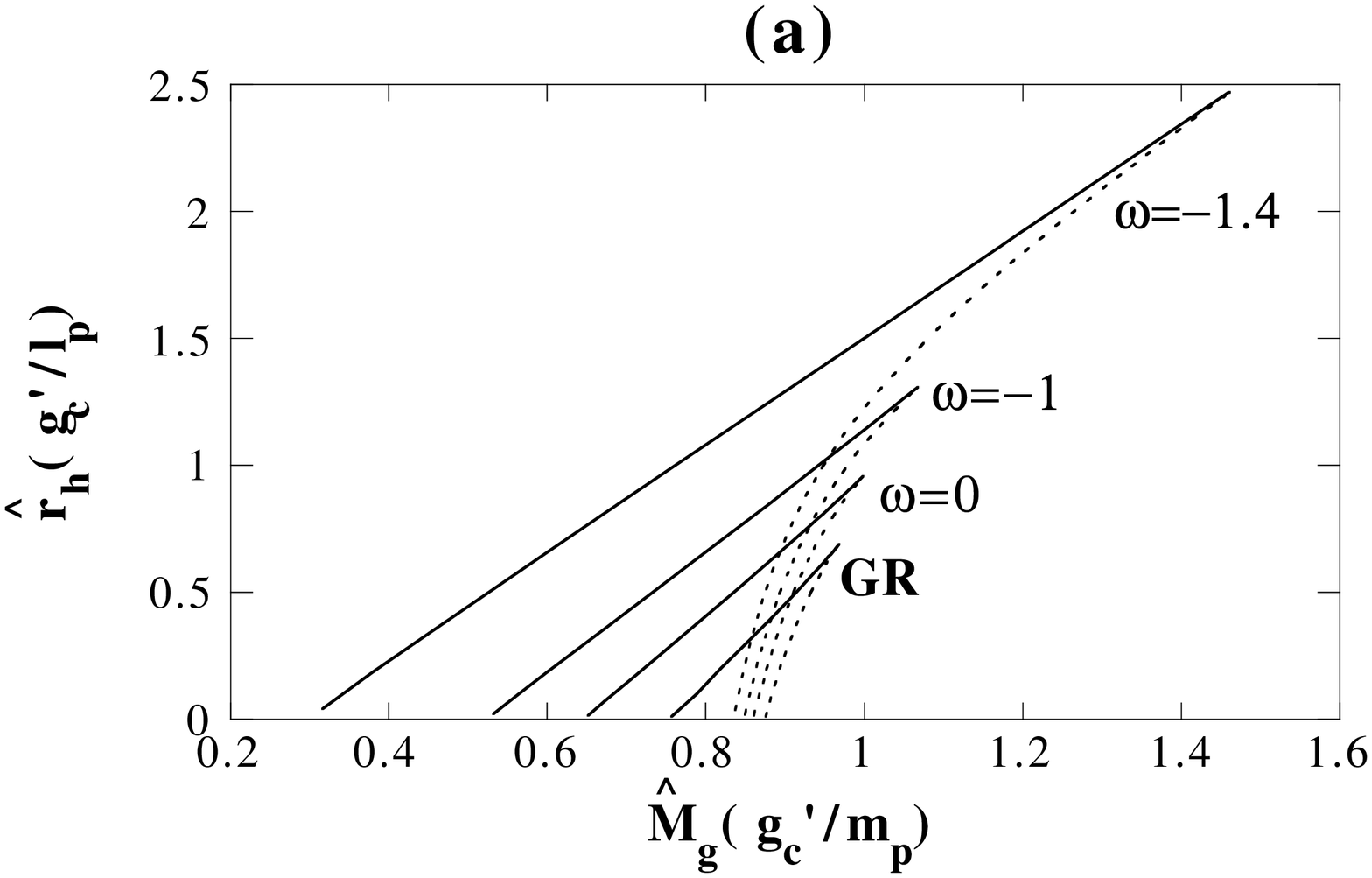}{}
\segmentfig{8cm}{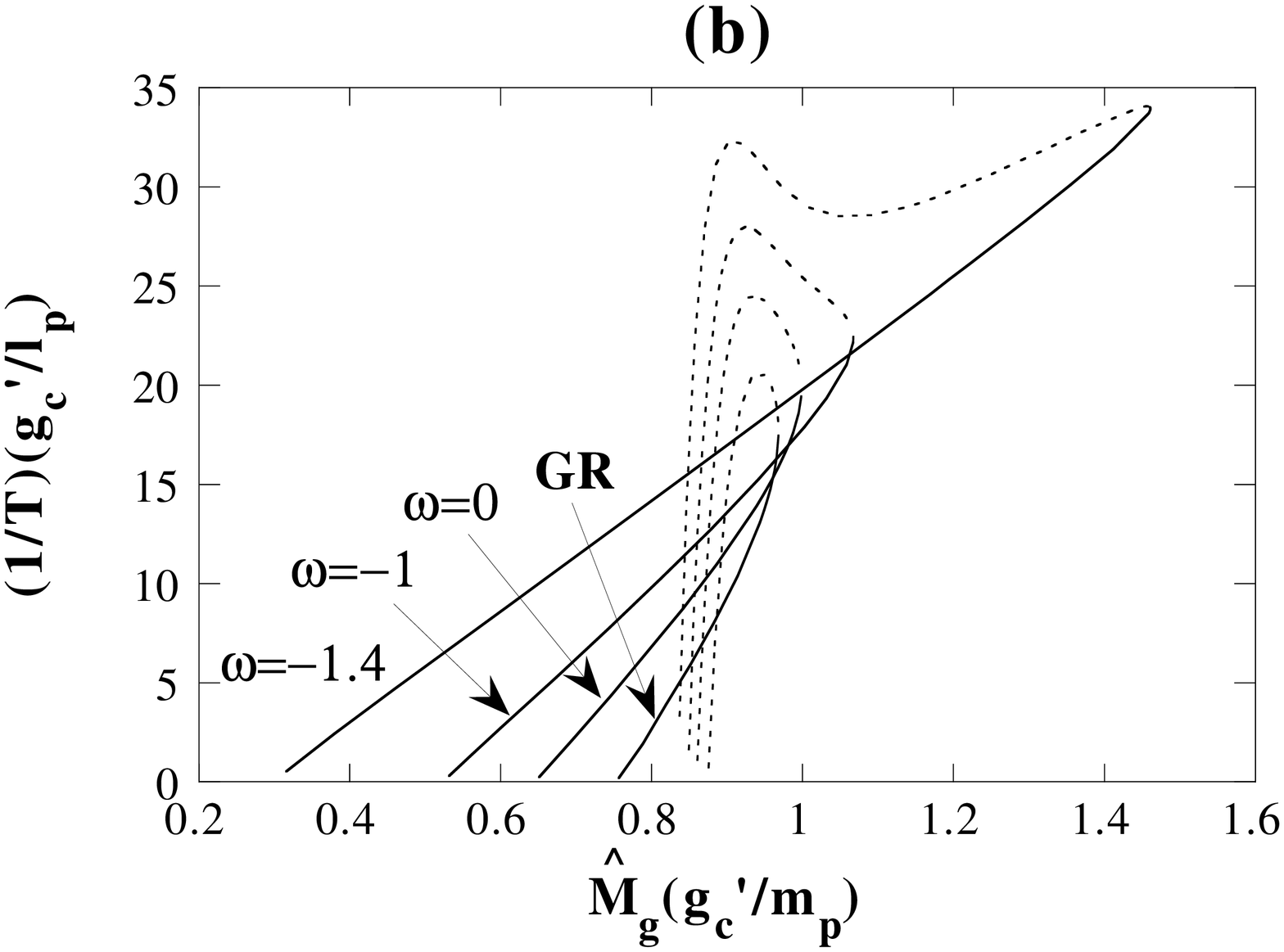}{}
\end{tabular}
\caption{
(a) $\hat{M}_{g}$-$r_{h}$ diagram and (b) $\hat{M}_{g}$-$1/T$ diagram  normalized by $g_{c}'=\protect\sqrt{\phi_{0}}$
$g_{c}$ 
in the Einstein frame for several values of $\omega$. The mass of the Proca field is 
$\mu=0.15 g_{c}m_{p}$.  \label{wdem-rPE2}
} 
\end{figure}

In order to see the difference between  BD
theory and GR more clearly, we show the
$\omega$ dependence of the black hole solutions in
Figs.\ref{wdem-rP},\ref{wdem-1TP}, and \ref{wdem-rPE2}.

From Fig.\ref{wdem-rPE2}, in which the effect
is absorbed in normalization by $g_{c}'$, we find the deviation
from GR is quite similar to the behavior when changing a mass of 
the Proca field in GR (see Fig.\ref{r-MBD}).  This means that effects 
(1) and (2) really explain the deviation from GR.

\begin{figure}[htbp]
\begin{tabular}{ll}
\segmentfig{8cm}{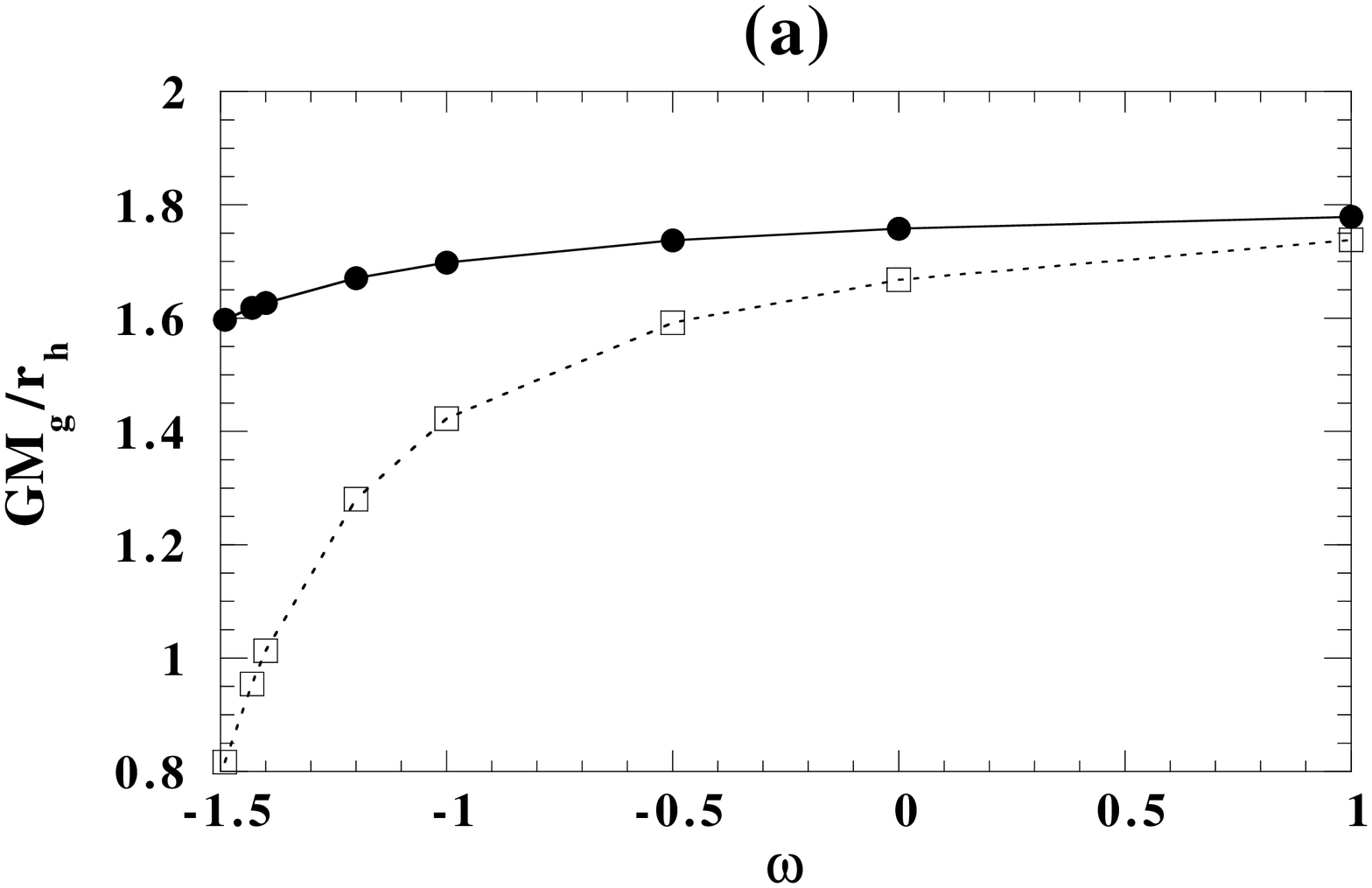}{}
\segmentfig{8cm}{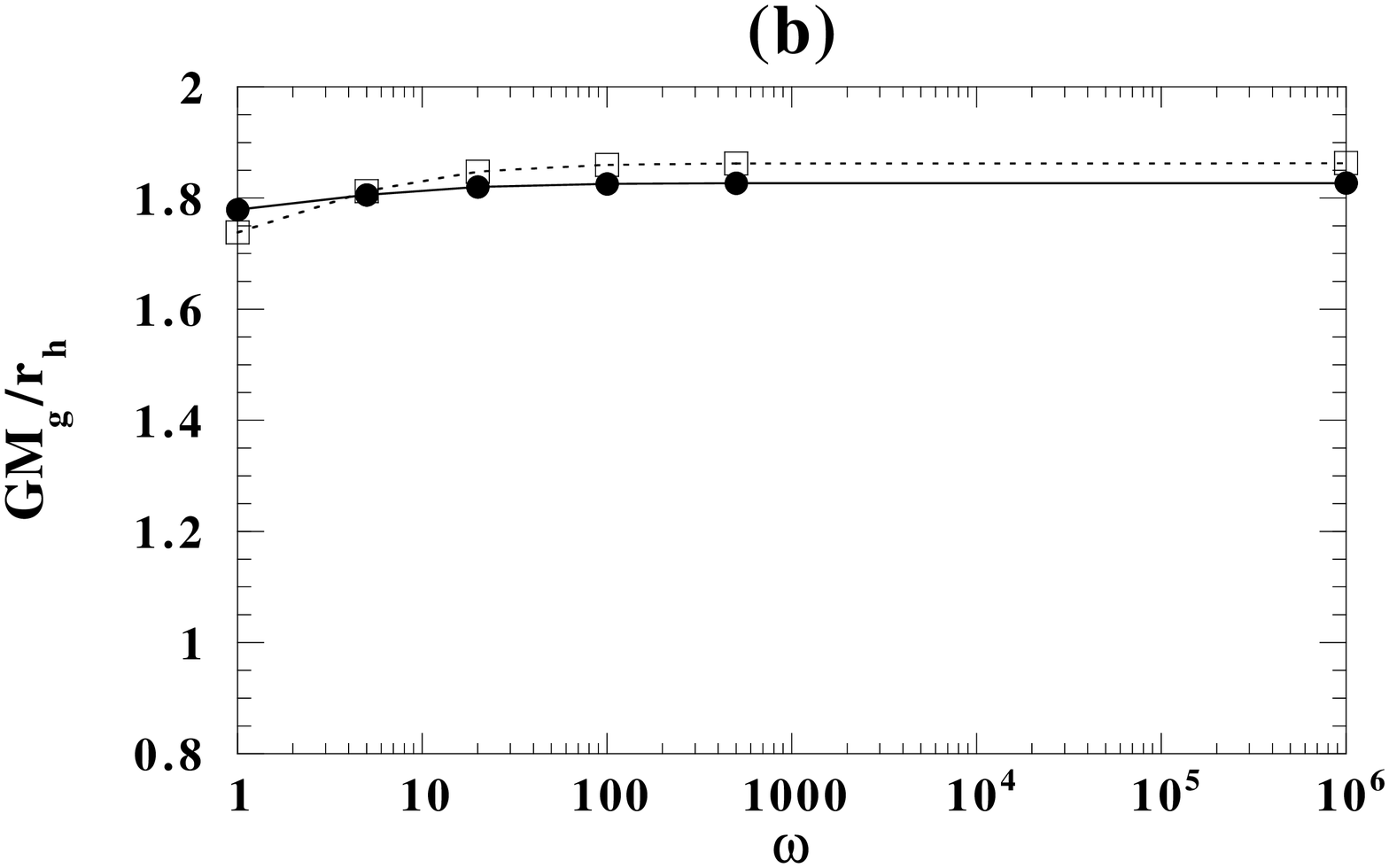}{}\\
\segmentfig{8cm}{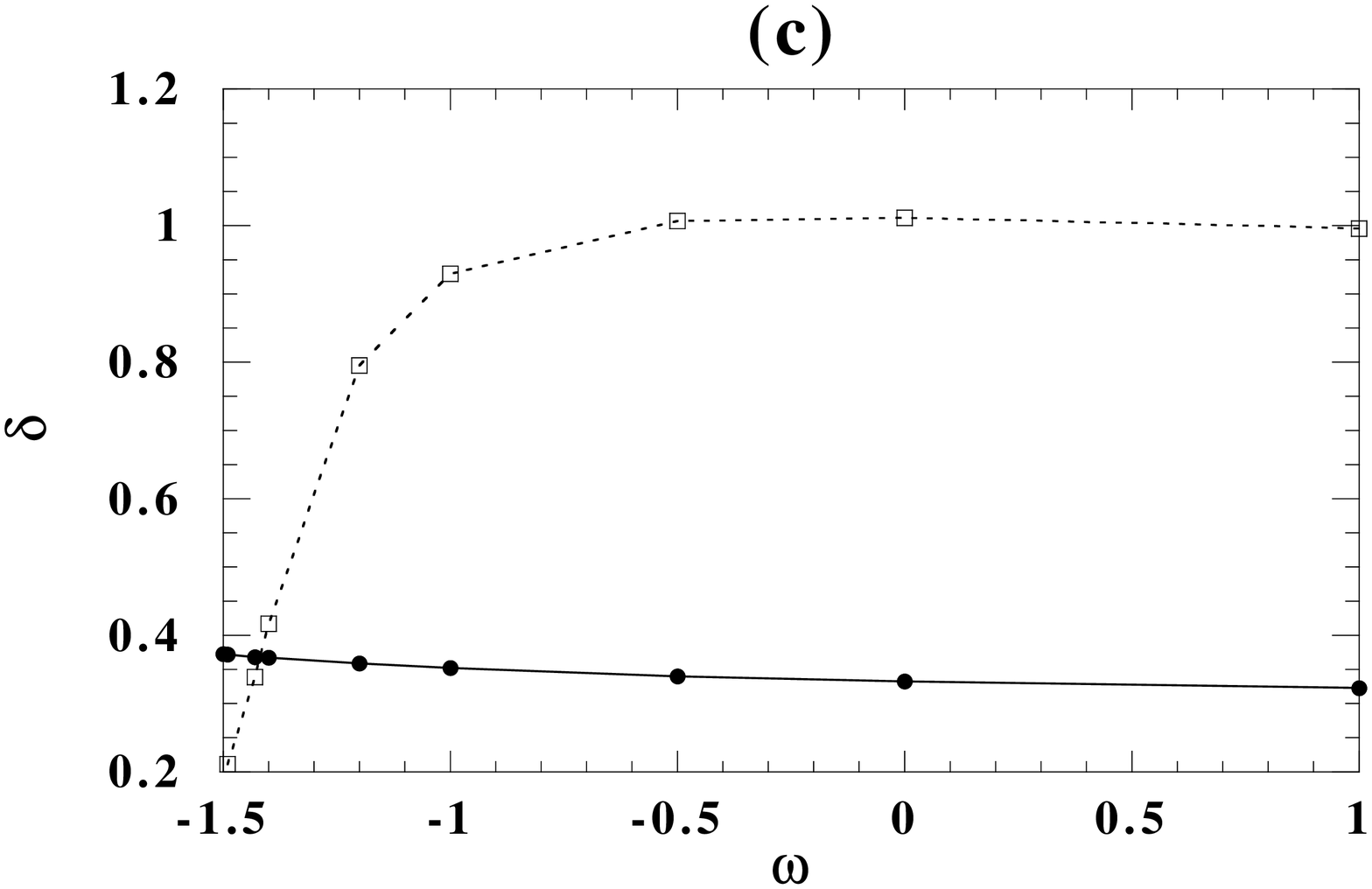}{}
\segmentfig{8cm}{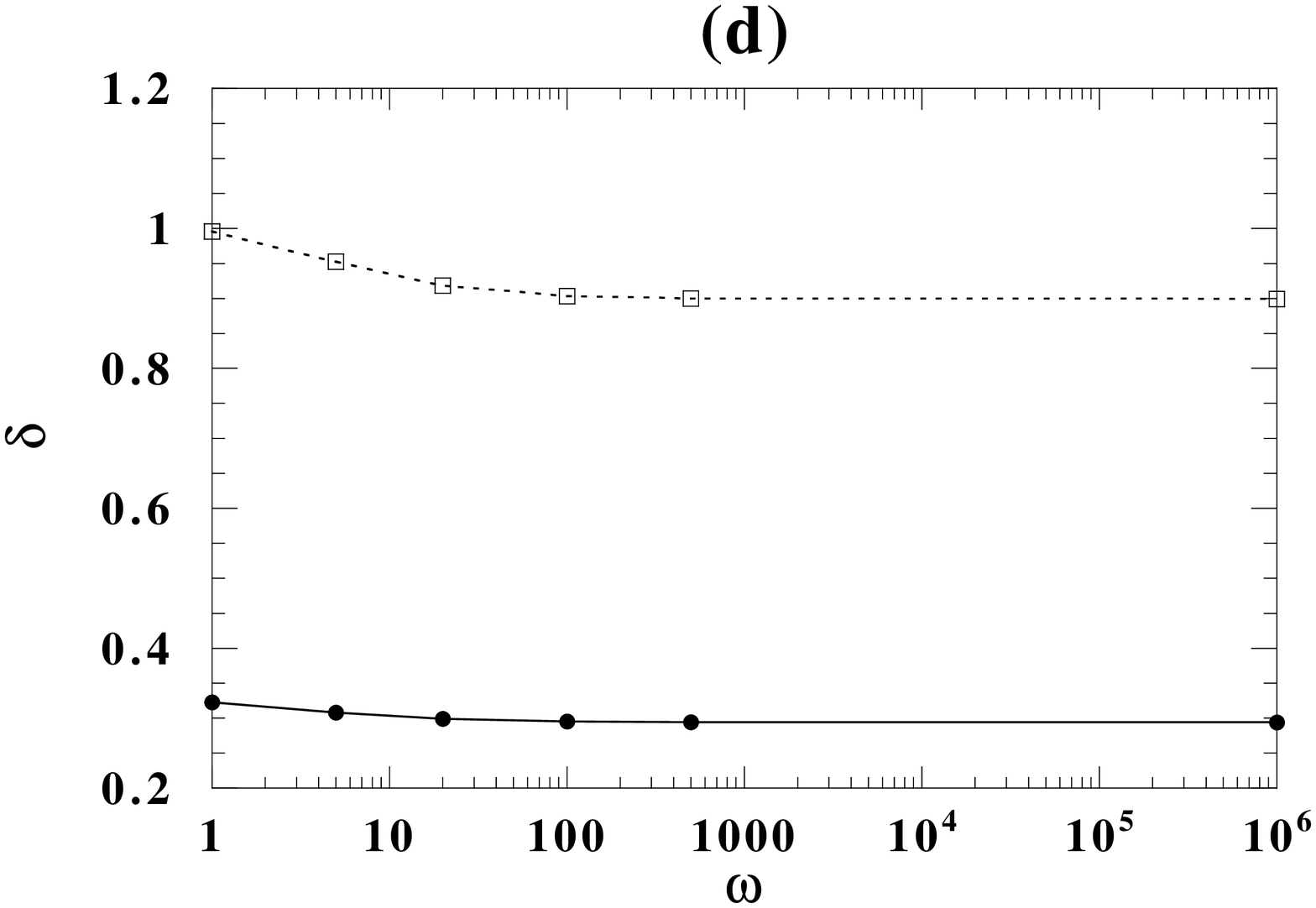}{}
\end{tabular}
\caption{$\omega$ dependence on the gravitational mass $M_{g}$
and the lapse function $\delta$ of the Proca black hole  in  the BD frame for a fixed
$r_{h} =0.5l_{p}/g_{c}$ and $\mu=0.15g_{c}m_{p}$ ((a) , (c) : 
$\omega \leq 1$ and (b) , (d) : $\omega \geq 1$). The dotted and solid lines correspond to the dotted- and solid-line branches 
in Fig.\protect\ref{wdem-rP}. \label{PMdel-W}   }
\end{figure}

In Fig.\ref{PMdel-W}, we depict the gravitational mass and the 
lapse function in
terms of $\omega$ for fixed horizon radius ($r_{h}=0.5l_{p}/g_{c}$) and 
fixed mass of the Proca field ($\mu=0.15g_{c}m_{p}$). The solid and
dotted lines correspond to those   in the  solid-line and   in
the dotted-line branches, respectively.   Note that, when we fix the horizon radius in the BD frame (or in the Einstein
frame),  the horizon radius in the Einstein frame (or in the BD
frame) will change for different values of $\omega$.
\begin{figure}[htbp]
\begin{tabular}{ll}
\segmentfig{8cm}{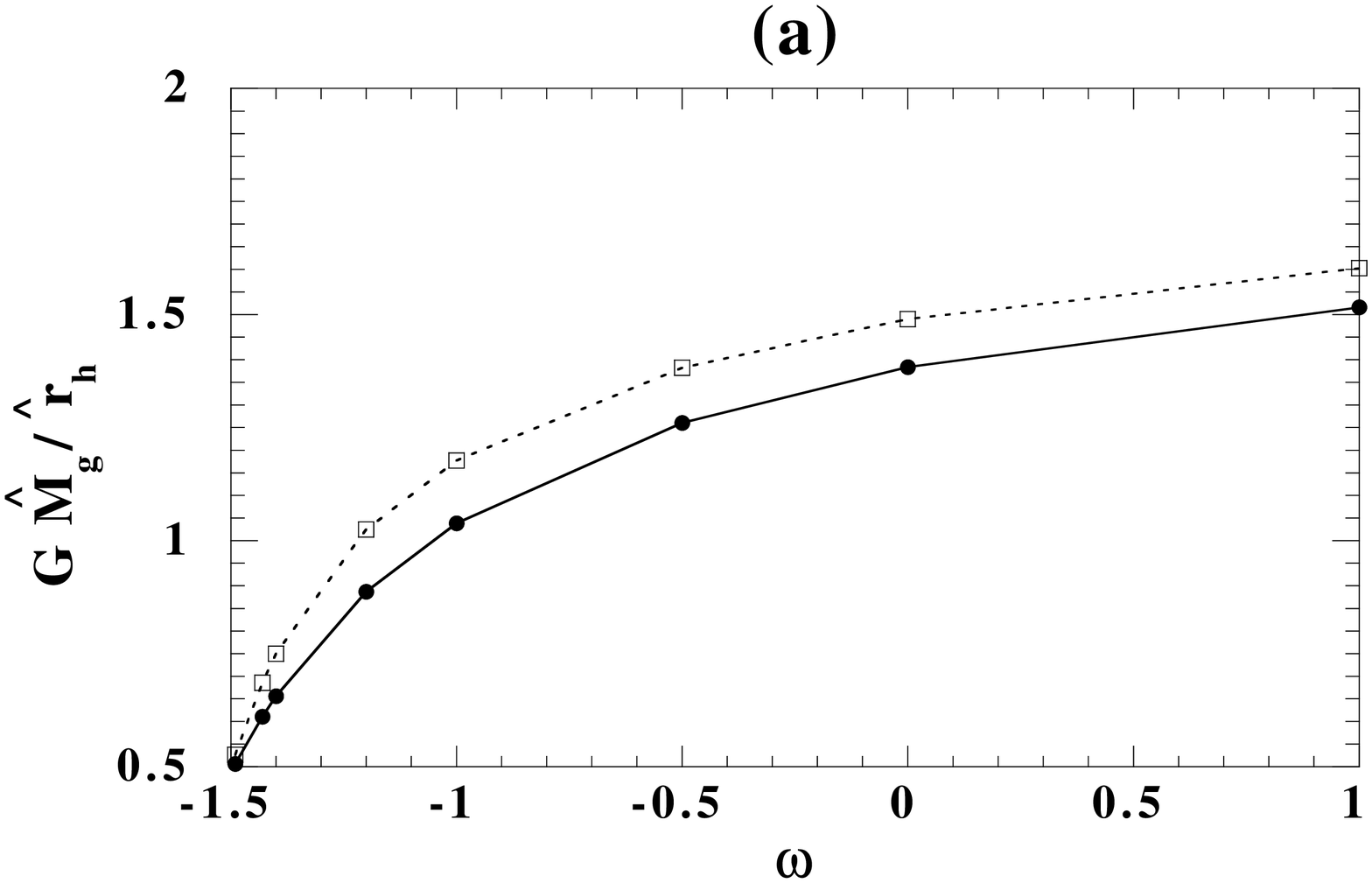}{}
\segmentfig{8cm}{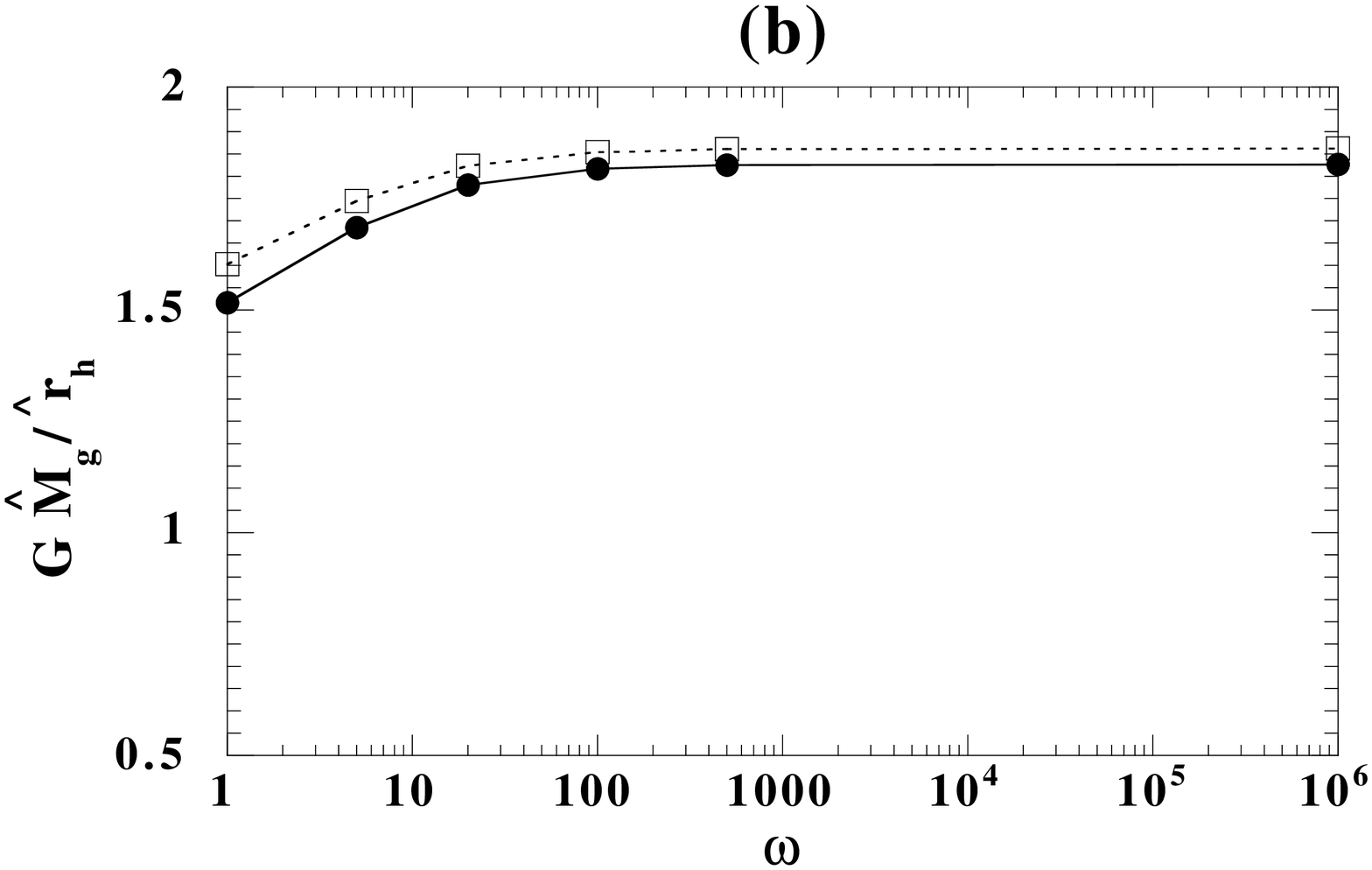}{}\\
\segmentfig{8cm}{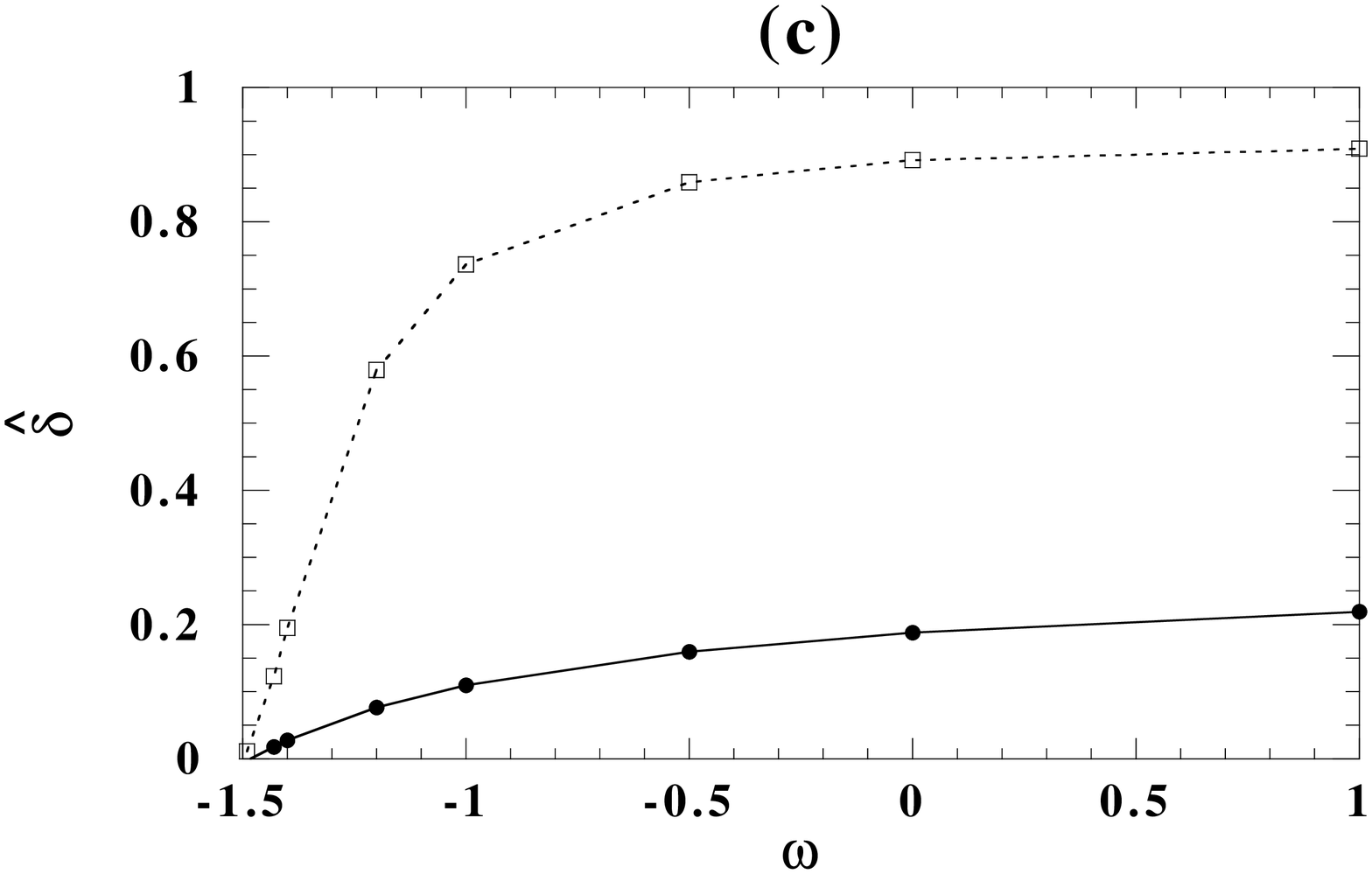}{}
\segmentfig{8cm}{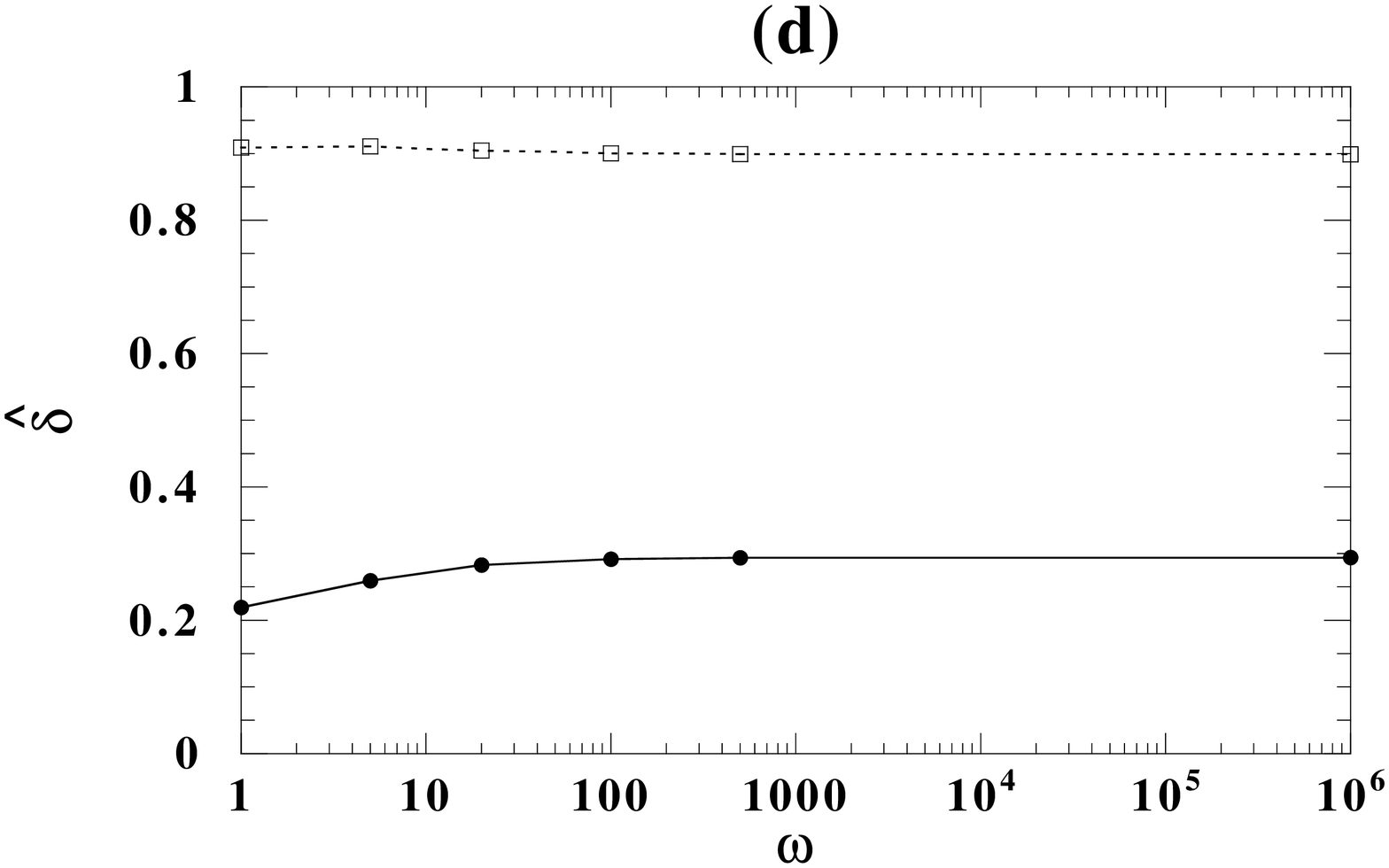}{}
\end{tabular}
\caption{$\omega$ dependence on the gravitational mass $\hat{M}_{g}$
and the lapse function $\hat{\delta}$ of the Proca black hole  in  the Einstein frame for a fixed $r_{h} =0.5l_{p}/g_{c}$ and $\mu=0.15g_{c}m_{p}$ ((a) , (c) : $\omega \leq 1$ and (b) , (d) : $\omega \geq 1$). The dotted and solid lines correspond to the dotted- and solid-line branches in 
Fig.\protect\ref{wdem-rP}.
\label{PMdel-Wefra}
}
\end{figure}
We can see that the gravitational mass approaches  some
constants as $\omega \rightarrow \infty$, which correspond to
those in GR. In the Einstein frame, the horizon radii in both
branches approach the Schwarzschild radius in the
limit of $\omega=-3/2$, resulting in a trivial Schwarzschild  
black hole(Fig.\ref{PMdel-Wefra}).

 This is because the matter contribution will
vanish in this limit as we discussed above ($\phi_0^{-1}
\rightarrow 0$).  In the BD frame, however, we find that the
dotted-line branch  changes faster than the solid-line branch and
 both horizon radii in the limit of $\omega=-3/2$ are different from the Schwarzschild radius. Then
 non-trivial black holes can exist even for $\omega=-3/2$.  This
is consistent with the above result in the Einstein frame because
the conformal transformation becomes singular for $\omega=-3/2$.

\begin{figure}[htbp]
\begin{tabular}{ll}
\segmentfig{8cm}{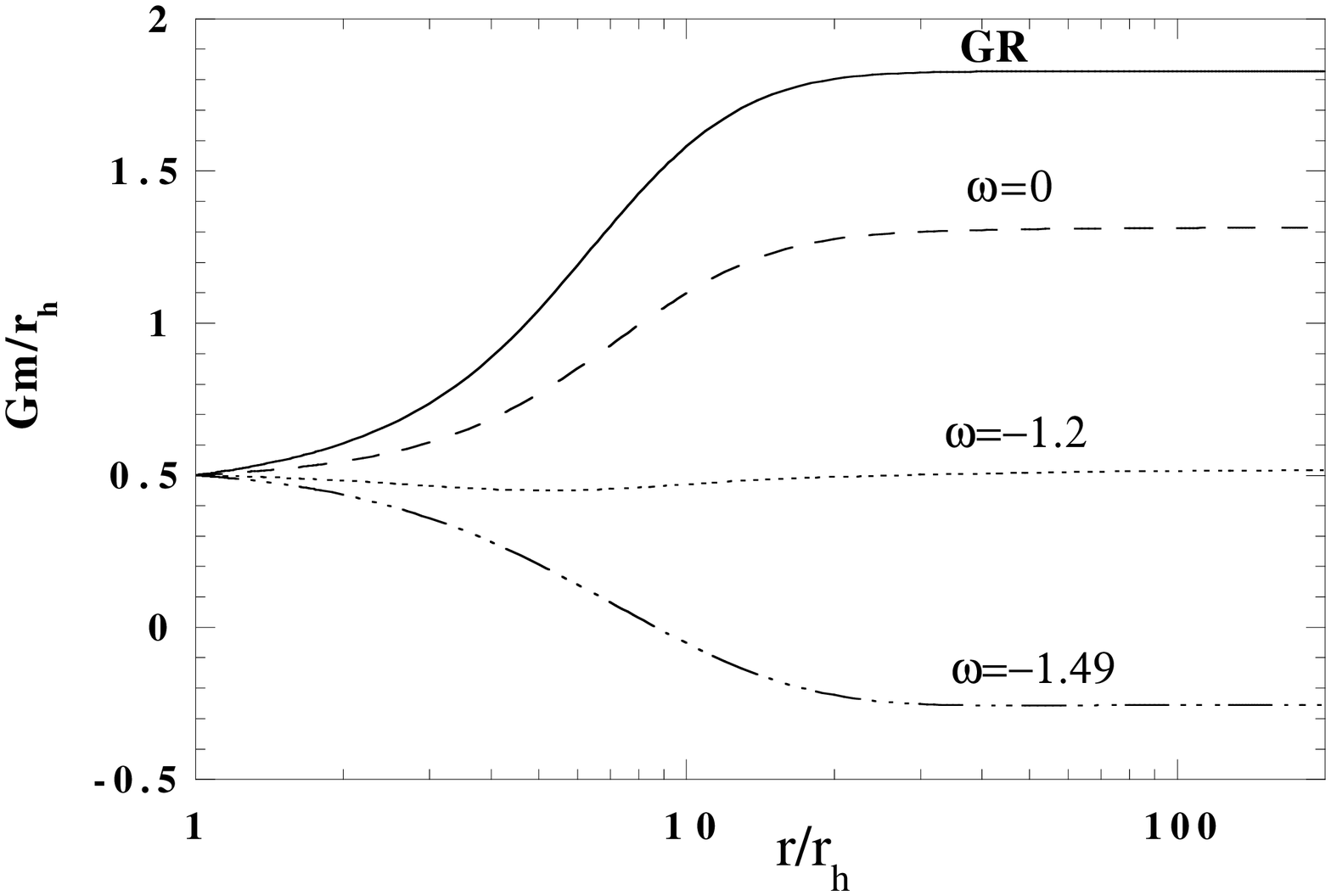}{(a)}
\hspace{5mm}
\segmentfig{8cm}{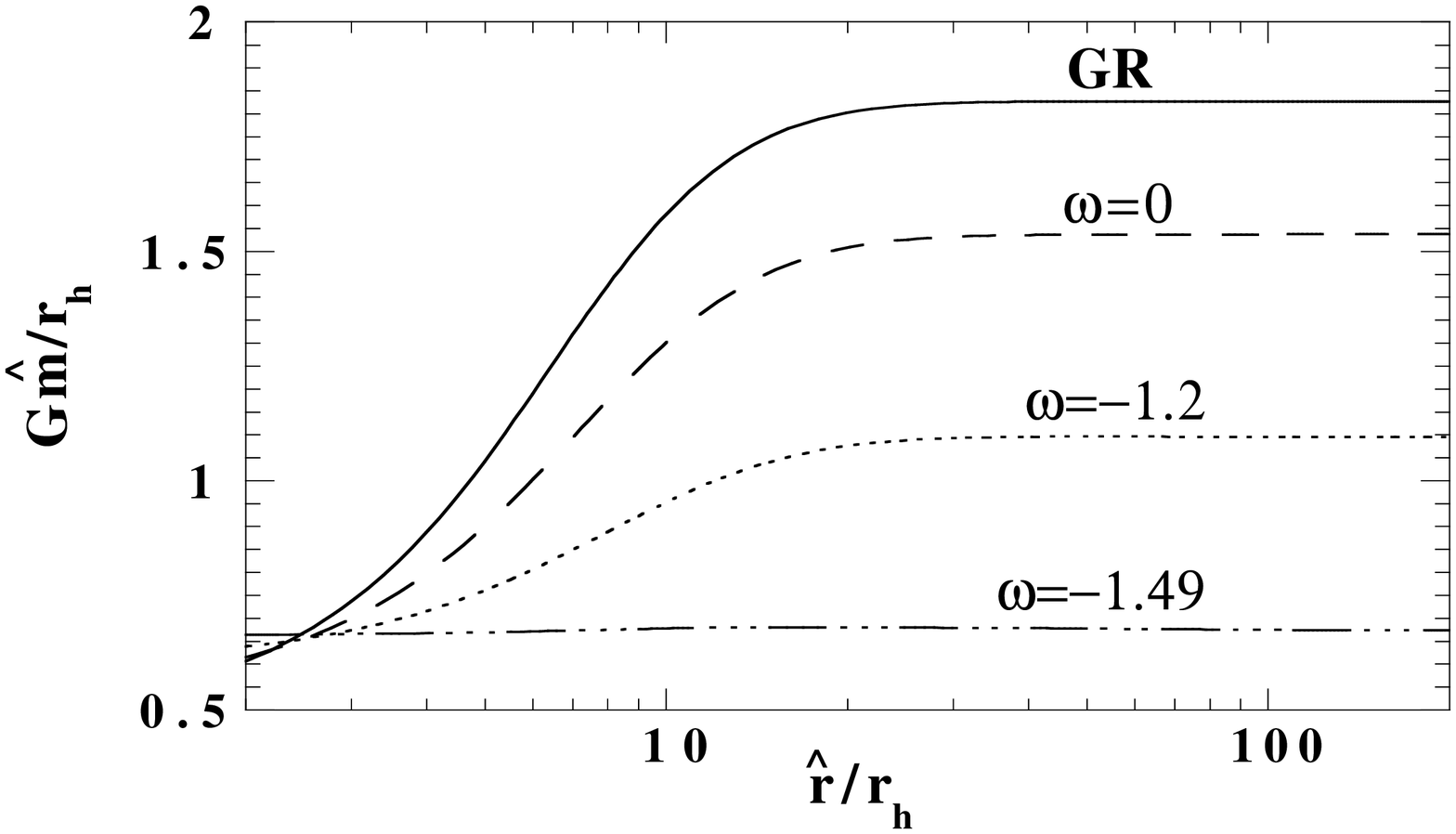}{(b)}
\end{tabular}
\vspace{5mm}
\caption{(a) The mass function $m(r)$ with
$r_{h}=0.5l_{p}/g_{c}$ and
$\mu=0.15g_{c}m_{p}$ in the BD frame for several values of $\omega$.
\label{r-mPBD} }
\vspace{5mm}
\caption{(b) The mass function $\hat{m}(\hat{r})$
 in the Einstein frame for several values of
$\omega$.
As in Fig. \protect\ref{r-mPBD}, we set  $\mu=0.15g_c m_p$ and
$r_{h}=0.5l_{p}/g_{c}$, which means that
$\hat{\lambda}_{h}$ is not fixed in this figure.
\label{r-mPE}}
\end{figure}

Although any value of $\omega > -3/2$ does not give a ghost, we
find a negative mass contribution in the BD frame, resulting in 
that $M$ becomes negative for $\omega <(\omega_{cr}<-1)$.  We show $m(r)$ for several values of $\omega$ in  Fig.\ref{r-mPBD}.  This does not mean, however, that
we have a negative-mass black hole, because the gravitational mass
$M_{g}$ is still positive. A test particle moving around a
black hole feels an attractive force given by $M_g$, which is always
positive. The effect of negative $M$ could be observed in a time
delay, which changes its sign for negative $M$\cite{Will}.

\begin{figure}
\begin{center}
\singlefig{8cm}{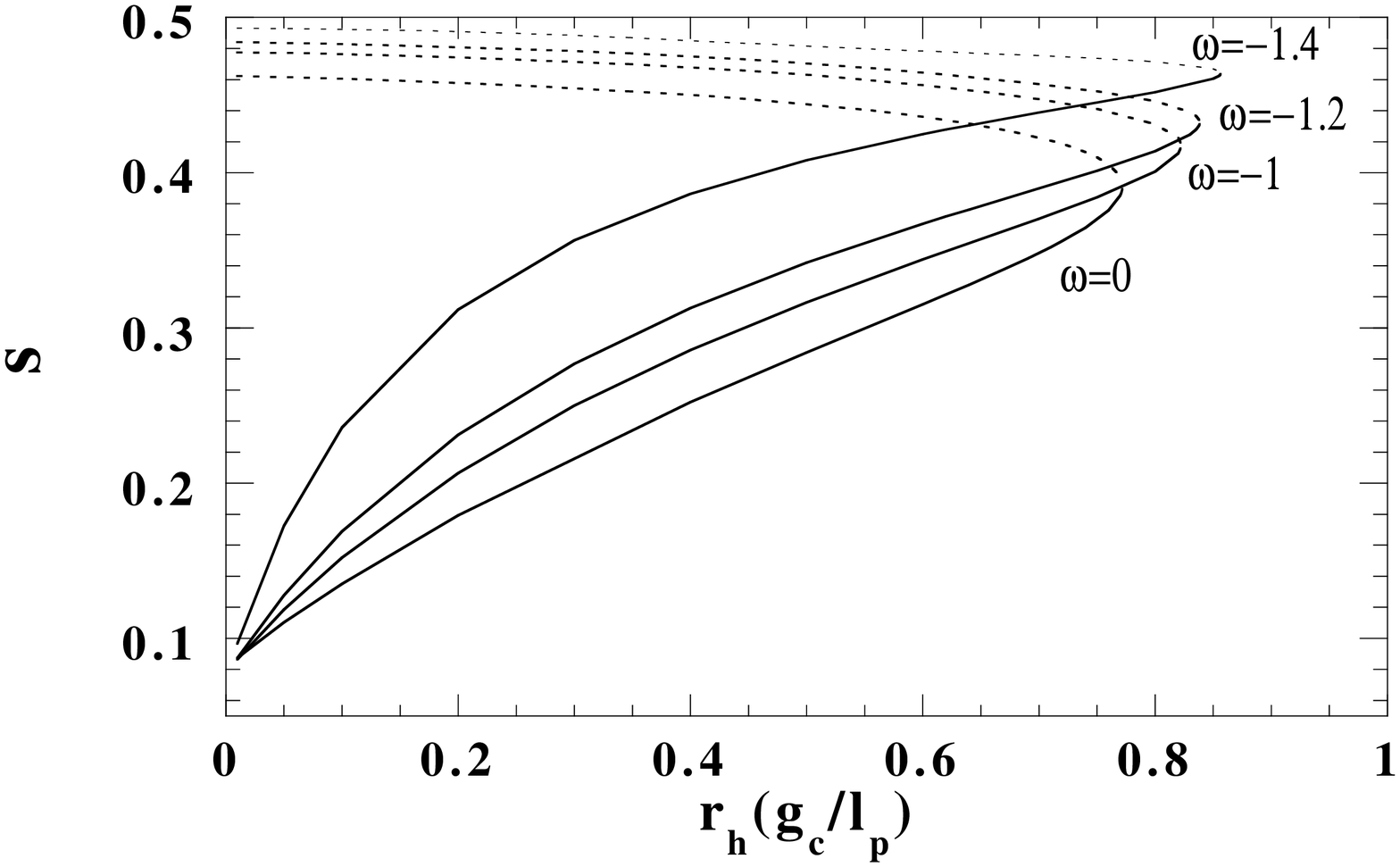}
\caption{$r_h$-$s$ diagram  for several values
of $\omega$. The mass of the Proca field is $\mu=0.15g_{c}m_{p}$. 
 The dotted and solid lines correspond to the dotted- and solid-line branches in  Fig.\protect\ref{wdem-rP}.
\label{sensitivity} }
\end{center}
\end{figure}

\begin{figure}[htbp]
\begin{tabular}{ll}
\segmentfig{8cm}{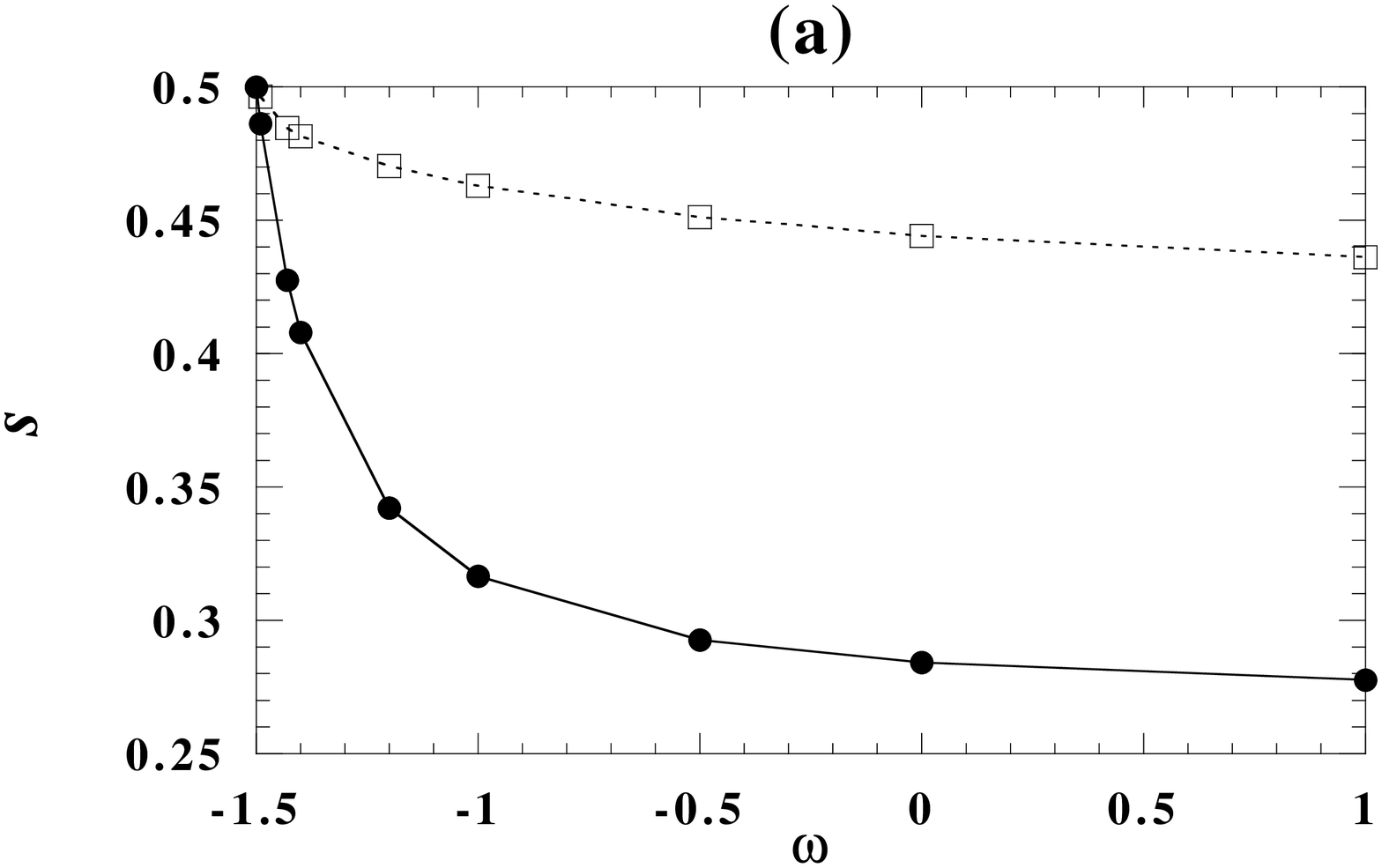}{}
\segmentfig{8cm}{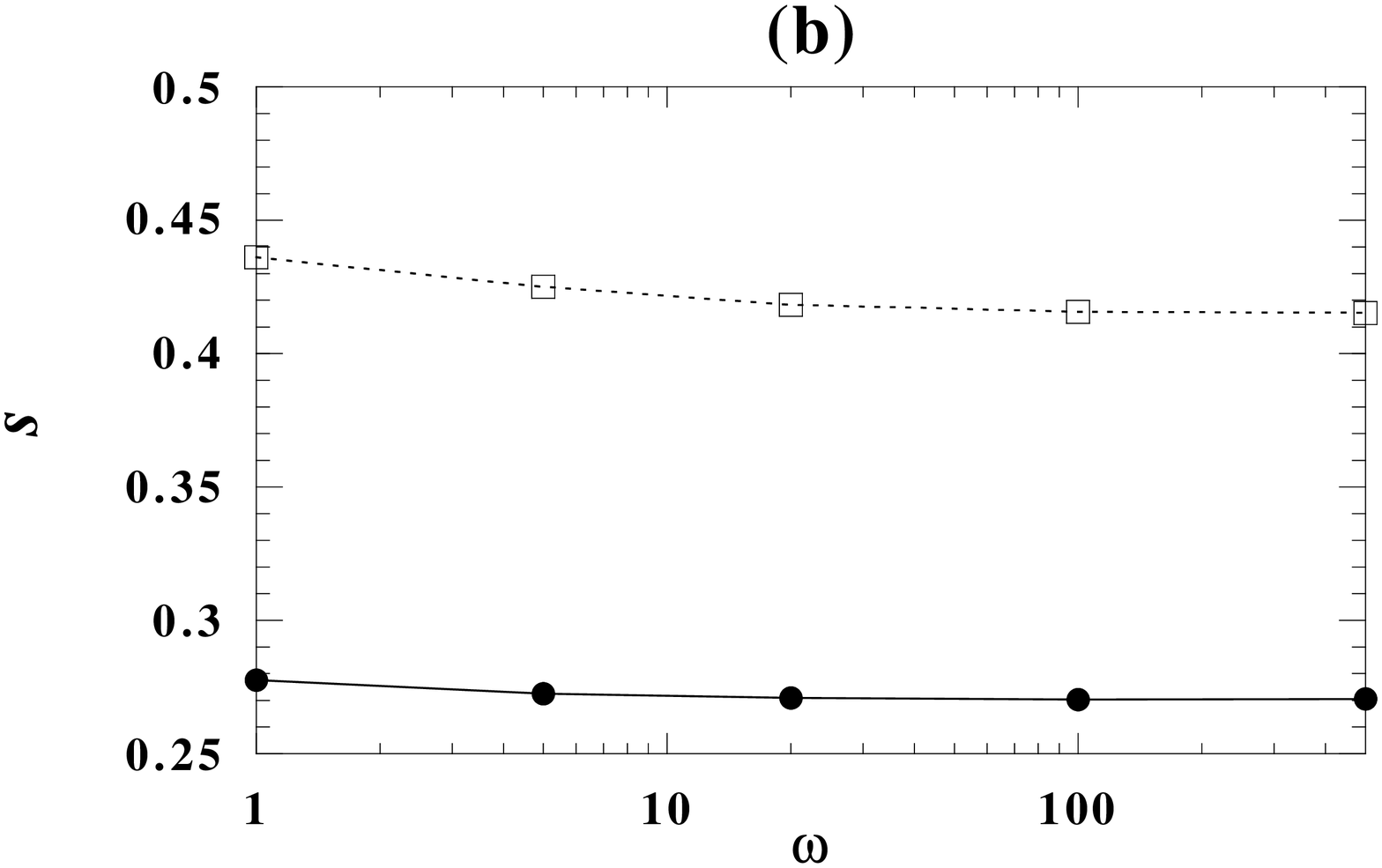}{}
\end{tabular}
\caption{$\omega$ dependence on the sensitivity $s$
 of the Proca black hole  for a fixed $r_{h} =0.5l_{p}/g_{c}$ and $\mu=0.15g_{c}m_{p}$ ((a) $\omega \leq 1$ and (b) $\omega \geq 1$). The dotted and solid lines correspond to the dotted- and solid-line branches in Fig.\protect\ref{wdem-rP}. \label{wdepsens}}
\end{figure}

In the Einstein frame, $\hat{m}(r)$ is monotonically increasing as
$r \rightarrow \infty$, resulting in a positive
mass $\hat{M}$, which is the same as the
gravitational mass  $\hat{M}_g$ (Fig.\ref{r-mPE}).  As we know, in
 BD theory, we can define several masses\cite{Will}.
The reason why we have several masses  is because the BD scalar
field decreases as $r^{-1}$, which is responsible for having
different masses in each frame,  and the scalar field itself also
gives a contribution into a mass energy as a scalar mass $M_s$.
In the vacuum case, we find a negative $M$ for
$\omega<-1$.  In our case, however, the BD scalar
field  is concentrated by the gravitational  attractive force of the black
hole. This changes the sensitivity $s$ just as for a self-gravitating
star.  From the asymptotic behavior of $g_{00}$ and $g_{rr}$, 
we find a relation between $M$ and $M_{g}$ as 
\begin{eqnarray}
\frac{ M_{g} }{ M }=\frac{ \omega+2-s }{ \omega+1+s }\ ,
\label{eq:mandmgbysens}
\end{eqnarray}
where $s$ is a sensitivity (see Eq.(11.83) in \cite{Will}).
 Then the sensitivity $s$ could be evaluated as
\begin{eqnarray}
 s=\frac{1}{2}-\frac{ (2\omega+3)(M_{g}-M) }{ 2(M+M_{g}) } \ .
\label{eq:sensitive}
\end{eqnarray}
For a Schwarzschild black hole ($M=M_{g}$), $s=1/2$. 
If $\omega=-3/2$, however, $s=1/2$ even if $M\neq M_{g}$. 
We show the sensitivity $s$ in Fig.\ref{sensitivity}. 
From Eq.(\ref{eq:mandmgbysens}), $M$ becomes negative 
for $\omega<-(1+s)\ (<-1)$. Then for a given $\omega\ (<-1)$, 
$M$ of the Proca black hole with smaller sensitivity than 
$s_{cr}\equiv -(1+\omega)$ becomes negative. It may correspond 
to smaller black holes in the solid-line branch 
from Fig.\ref{sensitivity}.   
When $\omega\rightarrow\infty$, we find $M=M_{g}$ but $s\neq 1/2$.  The reason is that the mass difference $M_{g}-M$(or $2M_{s}$) 
decreases as $\omega^{-1}$ for $\omega\rightarrow\infty$
(see Eq.(11.85) in \cite{Will}). 
While, as $\omega \rightarrow -3/2$,  $s\rightarrow 1/2$ 
but $M_{g}\neq M$ (or $M_{s}\neq 0$) (see Fig.\ref{wdepsens}). 
This is consistent with the previous fact that there still 
exists a nontrivial black hole in the limit of 
$\omega \rightarrow -3/2$.
%

%
\section{Skyrme Black Hole}                                       %
In GR, non-trivial black holes with a massive non-Abelian field
have quite similar properties, which we classified as Type II in
\cite{rf:Tachi}. How about black holes in BD theory?  To see whether 
the above results for the Proca black hole are generic  or not, we
shall  study  the Skyrme  field as another example
of a ``massive" non-Abelian field.

The action of  the Skyrme  field $L_{m}$ is
$SU(2)\times SU(2)$ invariant and given as\cite{Skyrme}
\begin{eqnarray}
 L_{m}=-\frac{1}{32 g_{s}^{2}}{\rm Tr}\mbox{\boldmath $F$}^{2}-
   \frac{f_{s}^{2}}{4}{\rm Tr}\mbox{\boldmath $A$}^{2}\ ,
\label{eq:s1}
\end{eqnarray}
where $f_{s}$ and $g_{s}$ are coupling constants. $g_{s}$ is
related to
$g_{c}$ for the Proca field as
\begin{eqnarray}
g_{s}=\sqrt{4\pi}g_{c}\ .  \label{eq:s1.5}
\end{eqnarray}
The ``mass" parameter of the Skyrme field,  $\mu$, is defined by
$\mu=f_{s}g_{s}$.
$\mbox{\boldmath $F$}$ and $\mbox{\boldmath $A$}$ are the  field
strength  and  its potential, respectively. They are  described by the $SU(2)$-valued function
$\mbox{\boldmath $U$}$ as
\begin{eqnarray}
\mbox{\boldmath $F$}= \mbox{\boldmath $A$}\wedge\mbox{\boldmath
$A$}, ~~~
\mbox{\boldmath $A$}=\mbox{\boldmath $U$}^{\dag}
\nabla\mbox{\boldmath $U$}
\ .
\label{eq:s2}
\end{eqnarray}
 In the spherically symmetric and static case, we can set
\mbox{\boldmath $U$},
\begin{eqnarray}
\mbox{\boldmath $U$}(\chi)=\cos \chi (r)+
  i\sin \chi (r) \mbox{\boldmath $\sigma$}_{i}\hat{r}^{i}\ ,
\label{eq:s3}
\end{eqnarray}
where $\mbox{\boldmath $\sigma$}_{i}$ and $\hat{r}^{i}$ are the Pauli spin matrices and a radial normal, respectively.
The boundary condition for the total field energy to be finite is
\begin{eqnarray}
\lim_{r\rightarrow \infty}\chi =0\ .    \label{eq:s4}
\end{eqnarray}
For simplicity, we solve the present system  in the Einstein frame.
The equivalent action $\hat{S} \equiv S/\phi_0$ is
\begin{eqnarray}
 \hat{S}=\displaystyle\int d^{4}x \sqrt{-\hat{g}}\left[
\frac{\hat{R}}{2\kappa^{2}}-
\frac{1}{2}\varphi_{,\alpha}\varphi^{,\alpha}
-{1 \over \phi_0}\left(\frac{1}{32 g_{s}^{2}}{\rm Tr}
\mbox{\boldmath $F$}^{2}+\frac{f_{s}^{2}}{4}
\exp{(-\kappa\beta\varphi)}{\rm Tr}\mbox{\boldmath
$A$}^{2}\right)\right]\ .
\label{eq:s5}
\end{eqnarray}
With the dimensionless parameter
$ \bar{f}_{s}\equiv f_{s}/m_{p}$, the basic equations are now
\begin{eqnarray}
\frac{  d\bar{\hat{m}}  }{  d\bar{\hat{r}}  }&=&2\pi
\left[
\bar{\hat{r}}^{2} \left(  1-\frac{  2\bar{\hat{m}}  }{
\bar{\hat{r}}  }
\right)
\left\{
\left(  \frac{  d\bar{\varphi}  }{  d\bar{\hat{r}}  } \right)^{2}
+\left(  \frac{d\chi}{d\bar{\hat{r}}}  \right)^{2}
\left(
\frac{\bar{f}_{s}^{2}}{\phi_{0}}\exp{
(-\sqrt{8\pi}\beta\bar{\varphi})  } +\frac{  \sin^{2}\chi  }{
2\pi\phi_{0}\hat{\lambda}_{h}^{2}\bar{\hat{r}}^{2}  }
\right)
\right\}  \right.  \nonumber  \\
&&\left.+\sin^{2}\chi
\left(
2\frac{\bar{f}_{s}^{2}}{\phi_{0}}\exp{
(-\sqrt{8\pi}\beta\bar{\varphi})  }+
\frac{  \sin^{2}\chi  }{
4\pi\phi_{0}\hat{\lambda}_{h}^{2}\bar{\hat{r}}^{2}   }
\right)
\right]   \ ,
\label{eq:sefra1}  \\
\frac{  d\hat{\delta}  }{  d\bar{\hat{r}} }&=&-4\pi\bar{\hat{r}}
\left\{
\left(  \frac{  d\bar{\varphi}  }{  d\bar{\hat{r}}  } \right)^{2}
+\left(  \frac{d\chi}{d\bar{\hat{r}}}  \right)^{2}
\left(
\frac{\bar{f}_{s}^{2}}{\phi_{0}} \exp{
(-\sqrt{8\pi}\beta\bar{\varphi})  } +\frac{  \sin^{2}\chi  }{
2\pi\phi_{0}\hat{\lambda}_{h}^{2}\bar{\hat{r}}^{2}   }
\right)
\right\}  \ ,  \label{eq:sefra2}  \\
\frac{  d^{2}\bar{\varphi}  }{  d\bar{\hat{r}}^{2}  }&=&
\left(  1-\frac{  2\bar{\hat{m}}  }{  \bar{\hat{r}}  }
\right)^{-1}
\left[
\frac{  d\bar{\varphi}  }{  d\bar{\hat{r}}  }
\left\{
\frac{  \sin^{2} \chi  }{  \bar{\hat{r}}  }
\left(
8\pi\frac{ \bar{f}_{s}^{2} }{ \phi_{0} }\exp{
(-\sqrt{8\pi}\beta\bar{\varphi})  }+
\frac{  \sin^{2}\chi  }{  \phi_{0}\hat{\lambda}_{h}^{2}
\bar{\hat{r}}^{2}  }
\right)
-\frac{  2\bar{\hat{m}}  }{  \bar{\hat{r}}^{2}  }
\right\}\right.  \nonumber  \\
&&\left.-\sqrt{8\pi}\beta \frac{\bar{f}_{s}^{2}}{\phi_{0}}
\exp{  (-\sqrt{8\pi}\beta\bar{\varphi})  }
\frac{  \sin^{2}\chi  }{  \bar{\hat{r}}^{2}  }
\right]  \nonumber  \\
&&-\frac{2}{\bar{\hat{r}}} \frac{d\bar{\varphi}}{d\bar{\hat{r}}}
-\frac{ \sqrt{8\pi}\beta\bar{f}_{s}^{2}  }{2\phi_{0}} \exp{
(-\sqrt{8\pi}\beta\bar{\varphi}})
\left( \frac{d\chi}{d\bar{\hat{r}}} \right)^{2}
\ ,  \label{eq:sefra3}  \\
\frac{d^{2}\chi}{d\bar{\hat{r}}^{2}}&=&
-\frac{  \exp{  (-\sqrt{8\pi}\beta\bar{\varphi})
}  }  {  4\pi\bar{\hat{r}}^{2}
\hat{\lambda}_{h}^{2} \bar{f}_{s}^{2}
\exp{  (-\sqrt{8\pi}\beta\bar{\varphi})  } +2\sin^{2}\chi }
\left(
8\pi\hat{\lambda}_{h}^{2} \bar{f}_{s}^{2}\bar{\hat{r}}
-4\pi\sqrt{8\pi}\beta\bar{\hat{r}}^{2}\hat{\lambda}_{h}^{2}
\bar{f}_{s}^{2}
\frac{d\bar{\varphi}}{d\bar{\hat{r}}}
+\frac{d\chi}{d\bar{\hat{r}}}\sin 2\chi
\right)\frac{d\chi}{d\bar{\hat{r}}}
\nonumber  \\
&&+\left( 1-\frac{2\bar{\hat{m}}}{\bar{\hat{r}}}  \right)^{-1}
\left[
\frac{  \sin 2\chi  }
{  4\pi\bar{\hat{r}}^{2}\hat{\lambda}_{h}^{2}\bar{f}_{s}^{2}
\exp{  (-\sqrt{8\pi}\beta\bar{\varphi})  } +2\sin^{2}\chi  }
\left(
4\pi\hat{\lambda}_{h}^{2}\bar{f}_{s}^{2}\exp{
(-\sqrt{8\pi}\beta\bar{\varphi})  } +\frac{  \sin^{2}\chi  }{
\bar{\hat{r}}^{2}  }
\right)  \right.
\nonumber  \\
&&+\left. \frac{d\chi}{d\bar{\hat{r}}}
\left\{
\frac{\sin^{2} \chi}{\bar{\hat{r}}}
\left(
8\pi\frac{ \bar{f}_{s}^{2} }{ \phi_{0} }\exp{
(-\sqrt{8\pi}\beta\bar{\varphi})  }+
\frac{  \sin^{2}\chi  }{  \phi_{0}\hat{\lambda}_{h}^{2}
\bar{\hat{r}}^{2}  }
\right)
-\frac{2\bar{\hat{m}}}{\bar{\hat{r}}^{2}}
\right\}
\right]
\ .  \label{eq:sefra4}
\end{eqnarray}
As in the case of the Proca black hole, the square brackets in
(\ref{eq:sefra3})  and (\ref{eq:sefra4}) must  vanish at
$r_h$ for the horizon to be regular. Hence
\begin{eqnarray}
\frac{  d\chi  }{  d\bar{r}  }\bigg|_{\bar{r}=1}&=&
-\frac{  \phi_{0}\hat{\lambda}_{h}^{2}\sin 2\chi_{h}(
4\pi\bar{f}_{s}^{2}\hat{\lambda}_{h}^{2} e^{
-\sqrt{8\pi}\beta\bar{\varphi}_{h}  }+\sin^{2}\chi_{h}  )  }
{  (  4\pi\bar{f}_{s}^{2}\hat{\lambda}_{h}^{2} e^{
-\sqrt{8\pi}\beta\bar{\varphi}_{h}  }+2\sin^{2}\chi_{h}  )
\{
\sin^{2}\chi_{h}(  8\pi\bar{f}_{s}^{2}\hat{\lambda}_{h}^{2}
e^{  -\sqrt{8\pi}\beta\bar{\varphi}_{h}  }+\sin^{2}\chi_{h}  )
-\phi_{0}\hat{\lambda}_{h}^{2}
\}  }
\ ,      \label{eq:sseisoku1}   \\
\frac{  d\bar{\varphi}  }{  d\bar{r}  }\bigg|_{\bar{r}=1}&=&
\frac{  \sqrt{8\pi}\beta\bar{f}_{s}^{2}\hat{\lambda}_{h}^{2}
e^{  -\sqrt{8\pi}\beta\bar{\varphi}_{h}  }\sin^{2}\chi_{h}  }
{  \sin^{2}\chi_{h}(  8\pi\bar{f}_{s}^{2}\hat{\lambda}_{h}^{2}
e^{  -\sqrt{8\pi}\beta\bar{\varphi}_{h} }+\sin^{2}\chi_{h}  )
-\phi_{0}\hat{\lambda}_{h}^{2} }\ ,      \label{eq:sseisoku2}
\end{eqnarray}
where $\chi_{h}\equiv \chi(r_h)$ and $\varphi_{h}\equiv
\varphi(r_h)$.
$\chi_{h}$ and $\bar{\varphi}_{h} (=\varphi_h/m_p)$ are
shooting parameters and should be determined iteratively so that
the boundary conditions  (\ref{eq:2.2}) and (\ref{eq:s4}) are
satisfied.

\begin{figure}[htbp]
\begin{tabular}{ll}
\segmentfig{8cm}{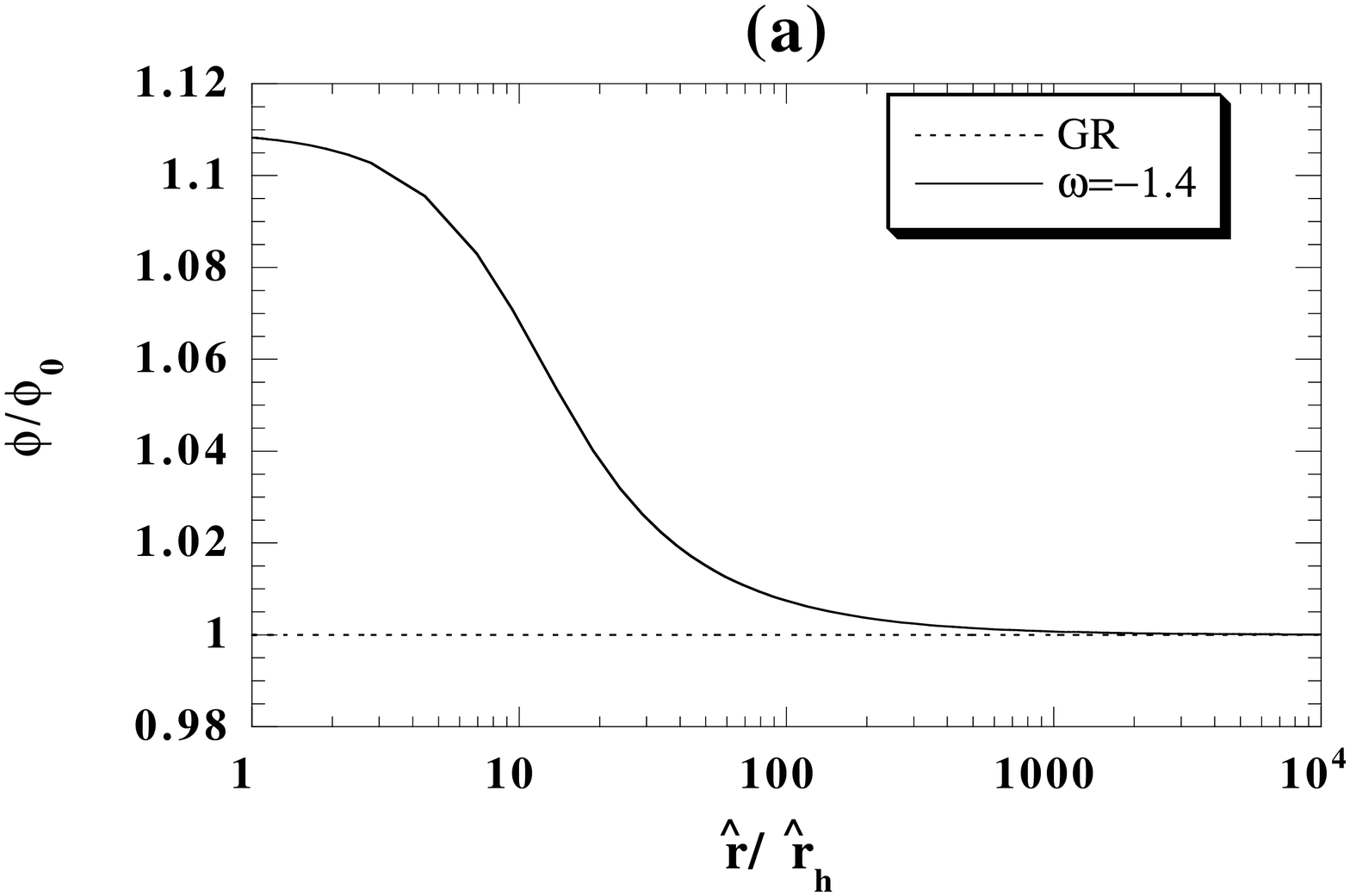}{}
\segmentfig{8cm}{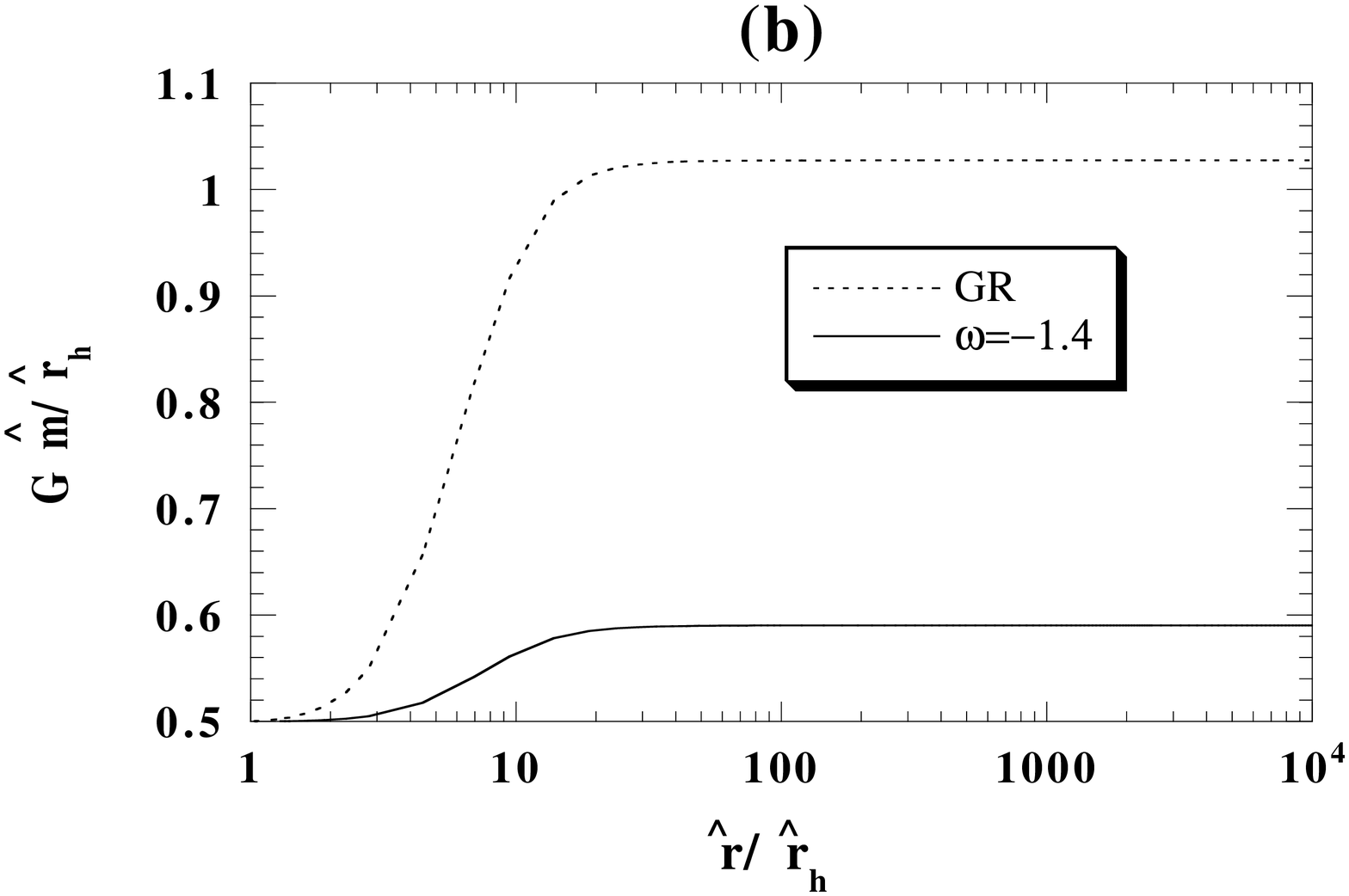}{}\\
\segmentfig{8cm}{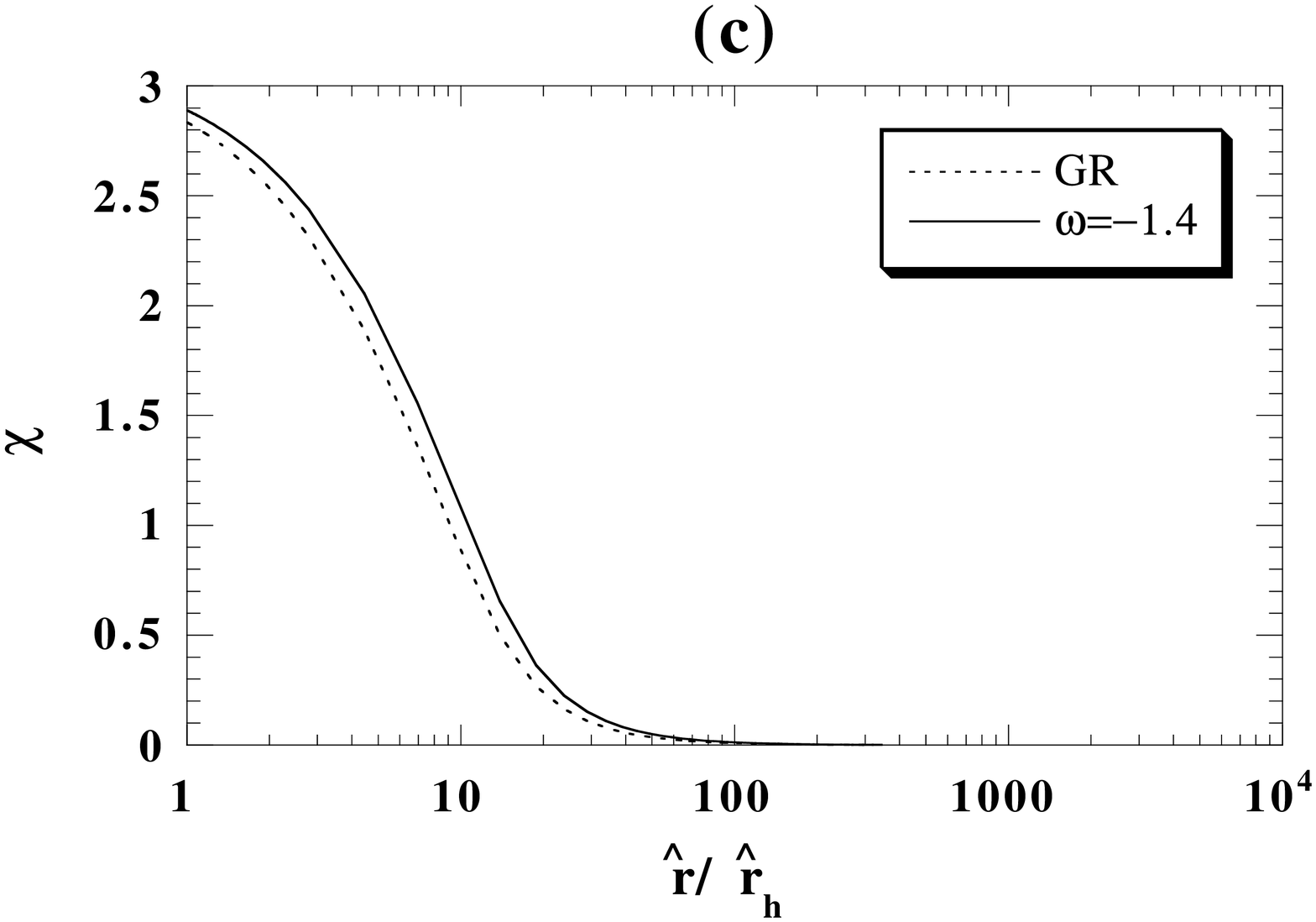}{}
\segmentfig{8cm}{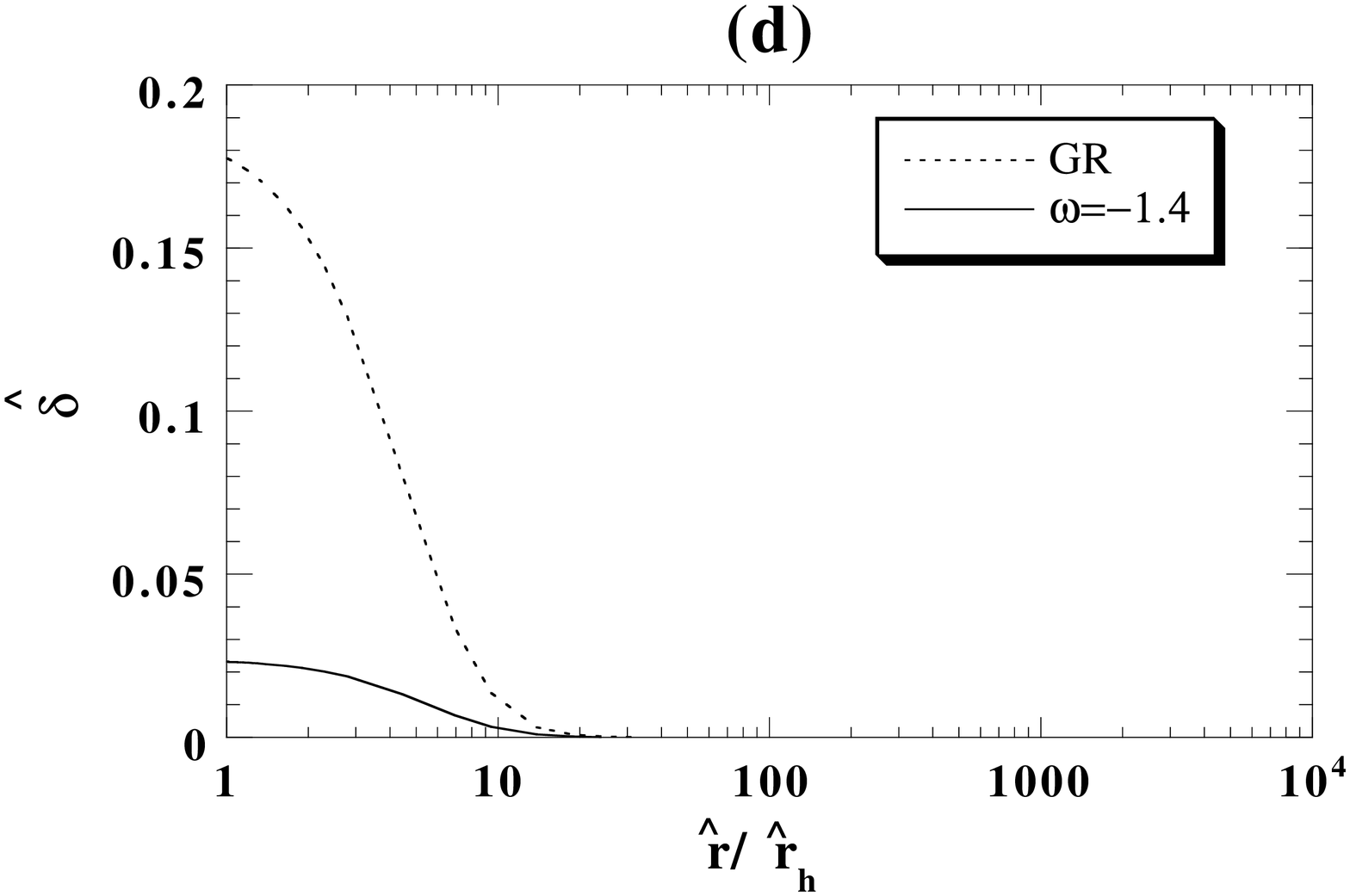}{}
\end{tabular}
  \caption{ The solution of the Skyrme black hole with
$\hat{r}_{h}= 1.0 l_p/g_c$ and $f_{s}=0.03 m_p$ in BD  theory  
( (a) $\phi (r)$, (b) $\hat{m}(r)$, (c) $\chi (r)$, (d) $\hat{\delta} (r)$ ). :  
The  Skyrme black hole in GR is also depicted as a reference.
\label{solution2}}
\end{figure}

We show a numerical result  of a Skyrme
black hole in BD theory in Fig.\ref{solution2}.

\begin{figure}[htbp]
\begin{tabular}{ll}
\segmentfig{8cm}{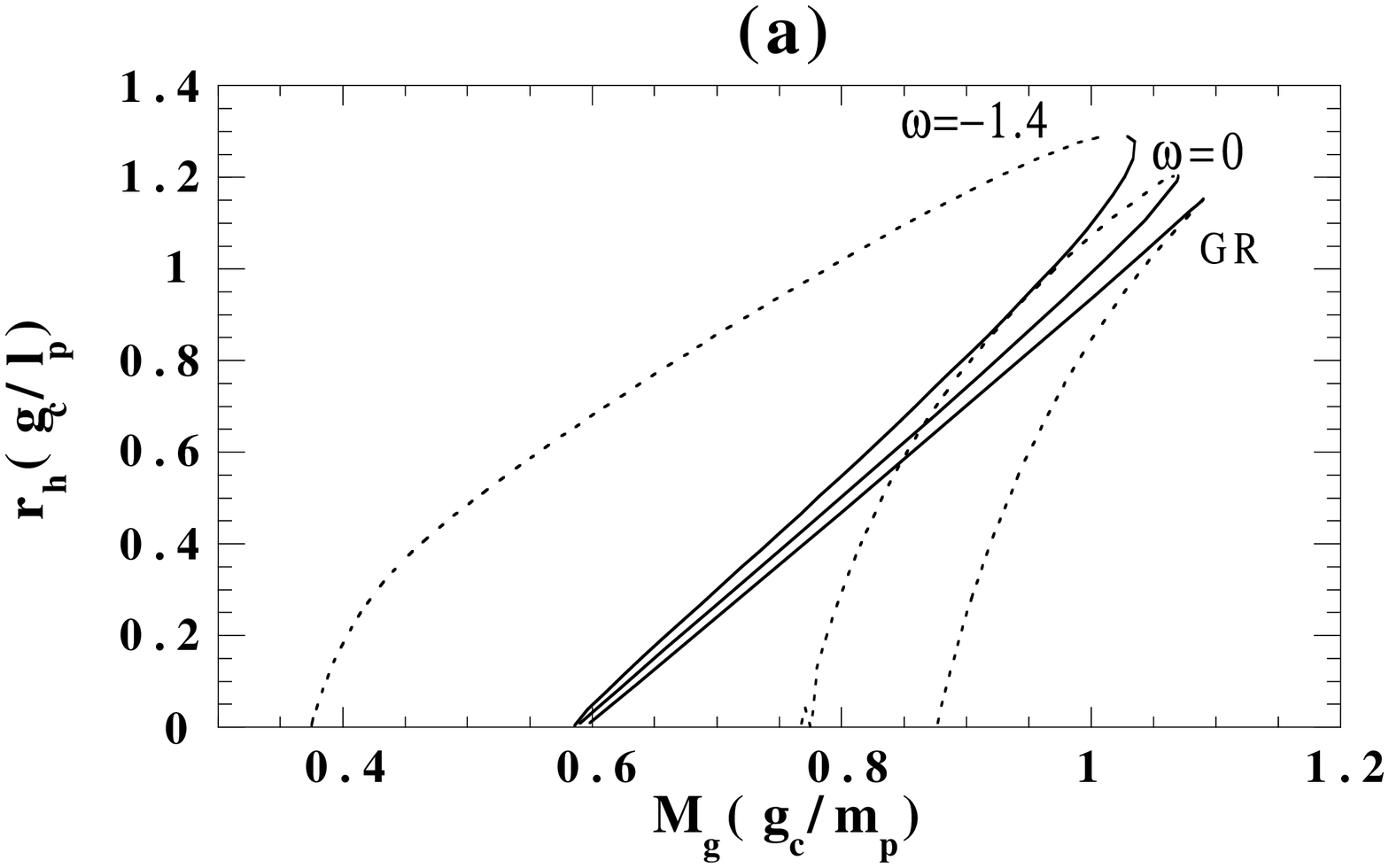}{}
\segmentfig{8cm}{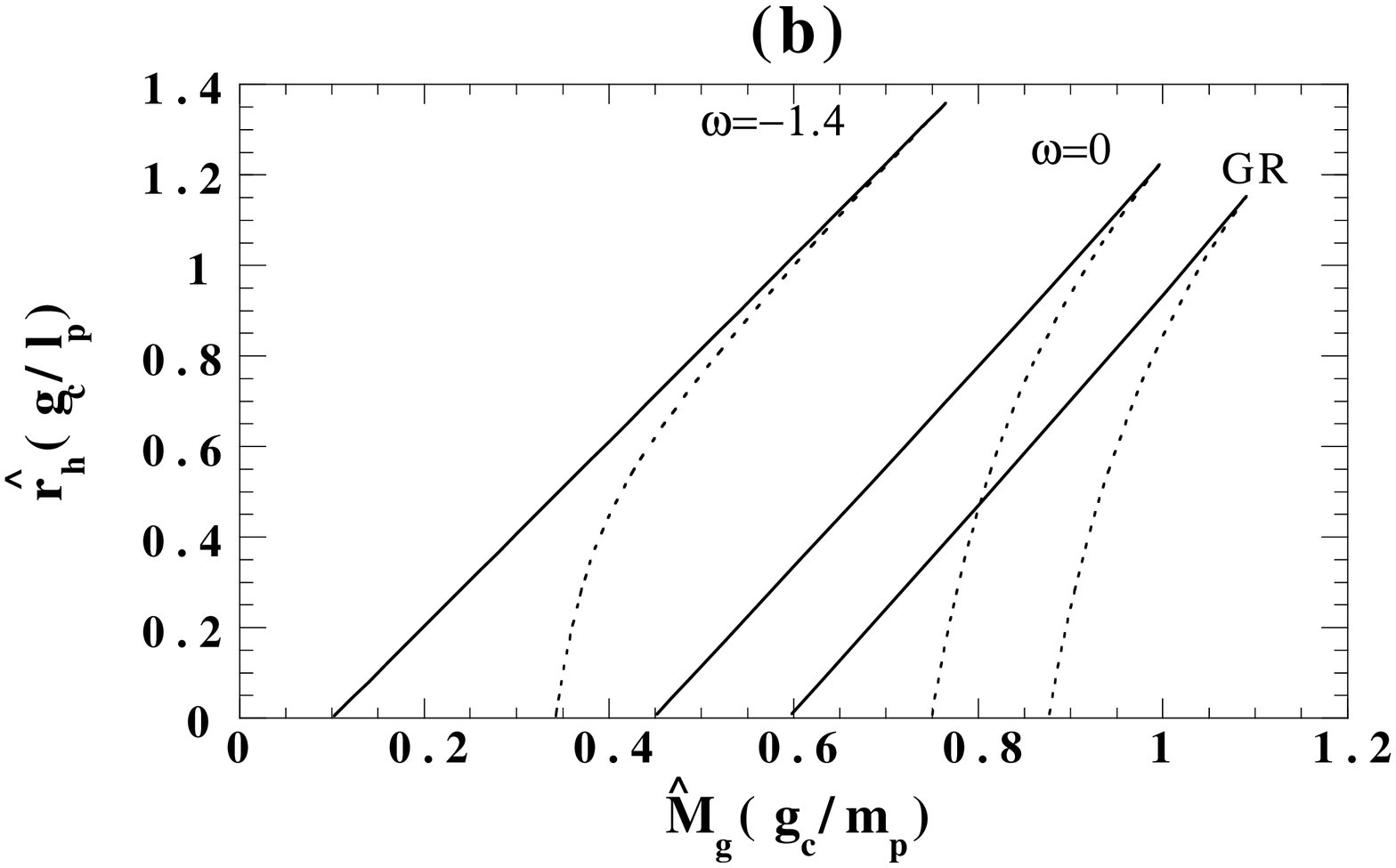}{}
\end{tabular}
\caption{(a)  $M_{g}$-$r_{h}$ diagram in the BD frame and (b)
 $\hat{M}_{g}$-$\hat{r}_{h}$ diagram in the Einstein
frame for Skyrme black holes.
We set  $f_s=0.03m_p$.
\label{SkyrmeM-R}}
\end{figure}

Here we set the parameters as
\begin{eqnarray}
\hat{r}_{h}= 1.0 l_p/g_c,
\ \ f_{s}=0.03 m_p,\ \ \omega=-1.4\ .
\label{eq:sbanofurumai}
\end{eqnarray}
 The dotted lines are those in GR with
$\hat{r}_{h}= 1.0 l_p/g_c$ and $f_{s}=0.03 m_p$.  The solutions 
correspond to solid-line in Fig.\ref{SkyrmeM-R}. We have shown
only for a solution with one node number.
 For a Skyrme black hole, rather than the
node number, the solution is characterized by the ``winding"
number defined by\footnote{ For a particlelike solution (Skyrmion), 
the value of $\chi$ at the origin must be $\pi n$, where $n$ is an 
integer and $|n|$ denotes the winding number of the Skyrmion. In the case of the black hole solution, it is topologically trivial. But $W_{n}$ 
defined by (\ref{eq:s6}) is close to $n$, so we shall also call it the ``winding" number. }
\begin{eqnarray}
W_{n}\equiv \frac{1}{\pi}|\chi_{h}-\chi (\infty)-\sin
(\chi_{h})|\ .
\label{eq:s6}
\end{eqnarray}

We show for a solution with the ``winding" number one.
Note that the comparison is made in the Einstein frame for a
fixed  $\hat{r}_{h}$, which does not mean the horizon radii
with different $\omega$  in the BD frame are  the same.

$\hat{\delta}$ falls faster than
$\hat{r}^{-1}$ because (\ref{eq:sefra2})  is
\begin{eqnarray}
\frac{d\hat{\delta}}{d\hat{r}} \sim -4\pi\hat{r}
\left(  \frac{d\varphi}{d\hat{r}}  \right)^{2}
\ ,   \label{eq:sefra2infinity}
\end{eqnarray}
as $\hat{r} \rightarrow \infty$ and $\varphi$ vanishes faster than
$\hat{r}^{-1}$ (see Figs.\ref{solution2}(a), (d)).
Then, as in the Proca black hole,  $\hat{M} = \hat{M}_{g}$.

\begin{figure}[htbp]
\begin{tabular}{ll}
\segmentfig{8cm}{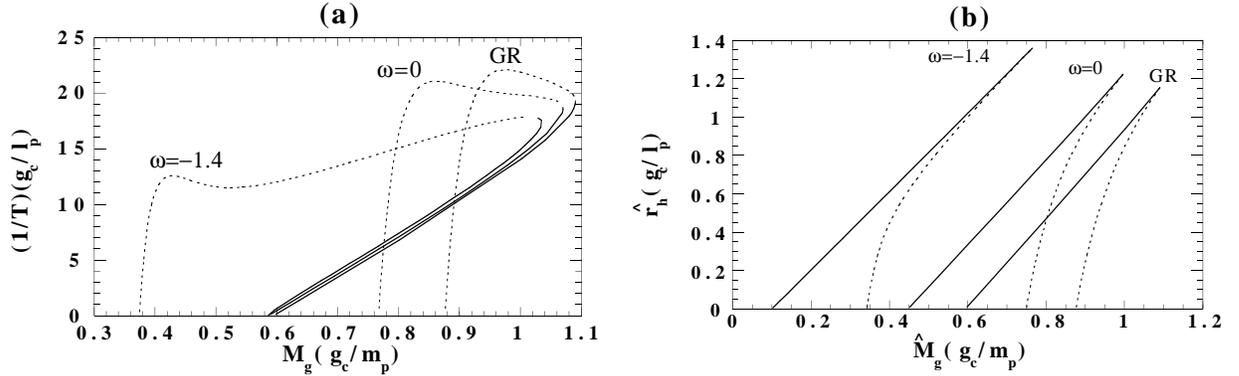}{}
\segmentfig{8cm}{SkyrmeEM-R.eps}{}
\end{tabular}
\caption{(a) $M_{g}$-$1/T$ diagram in the BD frame and
(b) $\hat{M}_{g}$-$1/T$ diagram in the Einstein frame for 
Skyrme black holes.
We set $f_s=0.03m_p$.\label{SkyrmeM-1T}
}
\end{figure}

To study the properties of a family of black holes,  we depict the
$\hat{M}_{g}$-$\hat{r}_{h}$   and $M_{g}$-$r_{h}$ (the BD
frame) diagrams in Fig.\ref{SkyrmeM-R} and the $M_{g}$-$1/T$
and $\hat{M}_{g}$-$1/T$ diagrams in Fig.\ref{SkyrmeM-1T}.
\newpage
We find that the results are quite similar to those for the  Proca
black holes. We have a  cusp structure in the $\hat{M}_{g}$-$\hat{r}_{h}$ diagram in
the Einstein frame, but it  disappears  in the BD frame.

Most properties found  for the Proca black hole apply
 to the Skyrme black hole as well.    This suggests that a
universal picture for non-trivial black holes with  massive
non-Abelian fields is possible.

\section{Concluding Remarks}          %
We have analyzed non-Abelian black holes (Proca and Skyrme
black holes) in BD theory  and shown  some  differences from
those in GR.  The Einstein conformal frame makes our analysis
easier.  The effect of the BD scalar field can be reduced into
two parts in the Einstein frame: the effective change of mass of
the non-Abelian field, i.e. $\mu \rightarrow \mu\exp
(-\kappa\beta\varphi/2)$ or $f_s \rightarrow f_s\exp
(-\kappa\beta\varphi/2)$, and the renormalized coupling
$g_c \rightarrow  \sqrt{\phi_0}g_c$ or $g_s \rightarrow
\sqrt{\phi_0} g_s$ and $f_s \rightarrow f_s/ \sqrt{\phi_0}$. As a
result, the solutions shift in the left-upper direction in the 
$M_g$-$r_h$ diagram.  Although we recover the Schwarzschild black
hole in the limit of $\omega \rightarrow -\frac{3}{2}$ in the Einstein
frame, we still have non-trivial black holes in the BD frame in the
same limit, because the conformal transformation becomes singular
then.

Secondly, we have analyzed for various values of $\omega$.
When $\omega \geq 500$, the difference from GR is so small that
we will not see any observational difference.  The
solutions for
$-\frac{3}{2} \leq
\omega \mbox{\raisebox{-1.ex}{$\stackrel
{\textstyle<}{\textstyle \sim}$}} -1$ seem to be somewhat  pathological, because the mass
function $m$ becomes negative in the BD frame, resulting in negative
value of
$M$.  However, even in such cases,
$M_{g}$  is always positive, therefore  a test particle around such
a black hole still  feels an attractive force.

Thirdly, we find that the cusp structure in  
$M_{g}$-$r_{h}$  diagram does not
appear in the BD theory although it was found in GR and provided us
a new method for stability analysis via catastrophe theory,
while  it exists in  the  Einstein frame.
 This suggests that a stability
change occurs at a cusp point in the Einstein frame.
The justification of this conjecture and the proper analysis
including that by linear perturbations will be given
elsewhere\cite{MTT}.

In this paper, we have studied a globally  neutral type of
non-Abelian black holes in BD theory.  A globally charged
black hole, i.e., a monopole black hole may be much more
interesting.  That is because charged black holes are important in the context of
cosmology, in particular, in the relation with a dynamical monopole
(topological inflation)\cite{VilenkinLinde}.
This is under investigation.

\section*{ACKOWLEDGEMENTS}
We would like to thank Jun-ichirou Koga for useful discussions 
and P. Haines for his critical reading.
T. Torii is  thankful for  financial support from the JSPS. This work
was supported partially by a Grant-in-Aid for  Scientific
Research Fund of the Ministry of Education, Science and Culture
(Specially Promoted Research No. 08102010 and No. 09410217), by a JSPS Grant-in-Aid
(No. 094162), and by the Waseda University Grant  for Special
Research Projects.

\end{document}